\documentclass[pra,aps,amsmath,amssymb,amsfonts,twocolumn,nofootinbib,longbibliography]{revtex4-1}
\usepackage{}
\usepackage{amssymb}
\usepackage{bm,mathrsfs}
\usepackage{graphicx}
\usepackage{epsfig}
\usepackage{amsmath,bbm}
\usepackage{amsfonts,amssymb}
\usepackage{times}
\usepackage{verbatim}
\usepackage[sort&compress]{natbib}
\usepackage{amsmath}
\usepackage{bm}
\usepackage{float}
\allowdisplaybreaks[4]
\usepackage[colorlinks,breaklinks,linkcolor=blue,anchorcolor=blue,citecolor=blue,urlcolor=magenta,dvipdfm]{hyperref}

\usepackage{color}
\definecolor{MyDarkBlue}{rgb}{0,0.08,0.45}
\definecolor{yellow}{rgb}{0.99,0.99,0.70}
\definecolor{white}{rgb}{1.0,1.0,1.0}
\definecolor{black}{rgb}{0.00,0.00,0.00}
\definecolor{green}{rgb}{0.8,0.98,0.83}
\definecolor{cGreen}{RGB}{0,230,0}
\definecolor{dkgreen}{rgb}{0,0.6,0}
\pagecolor{green}
\definecolor{zzz}{rgb}{0.9,0.0,0.4}
\pagecolor{white}

\begin{document}
\title{Non-Markovian dynamics with a giant atom coupled to a semi-infinite photonic waveguide}
\author{Z. Y. Li$^{1}$ and H. Z. Shen$^{1,2,}$}
\email{Corresponding author: shenhz458@nenu.edu.cn\\https://orcid.org/0000-0002-4017-7367}

\affiliation{$^1$Center for Quantum Sciences and School of Physics,
Northeast Normal University, Changchun 130024, China\\
$^2$Center for Advanced Optoelectronic Functional  Materials
Research, and Key Laboratory for UV Light-Emitting Materials and
Technology of Ministry of Education, Northeast Normal  University,
Changchun 130024, China}
\date{\today}

\begin{abstract}
We study the non-Markovian dynamics of a two-level giant atom interacting with a one-dimensional semi-infinite waveguide through multiple coupling points, where a perfect mirror is located at the endpoint of the waveguide. The system enters a non-Markovian process when the travel time of the photon between adjacent coupling points is sufficiently large compared to the inverse of the bare relaxation rate of the giant atom. The photon released by the spontaneous emission of the atom transfers between multiple coupling points through the waveguide or is reabsorbed by the atom with the photon emitted via the atom having completed the round trip after reflection of the mirror, which leads to the photon being trapped and forming bound states. We find that three different types of bound states can be formed in the system, containing the static bound states with no inversion of population, the periodic equal amplitude oscillation with two bound states, and the periodic non-equal amplitude oscillation with three bound states. The physical origins of three bound states formation are revealed. Moreover, we consider the influences of the dissipation of unwanted modes and dephasing on the bound states. Finally, we extend the system to a more general case involving many giant atoms coupled into a one-dimensional semi-infinite waveguide. The obtained set of delay differential equations for the giant atoms might open a way to better understand the non-Markovian dynamics of many giant atoms coupled to a semi-infinite waveguide.
\end{abstract}



\maketitle
\section{Introduction}
The core topic of quantum optics \cite{Kockum192019} is the understanding and application of the interaction between light and matter. The study of the interaction between light and matter can be well carried out on platforms such as cavity quantum electrodynamics (QED) systems \cite{Kimble1271998,Raimond5652001,Walther13252006}, circuit-QED systems \cite{Blais0250052021,Blais0623202004,Wallraff1622004}, and waveguide QED systems \cite{Roy0210012017,Roy12017,Sheremet210306824}, the limited bandwidth of the waveguide QED systems can be relaxed \cite{Roy0210012017} because the waveguides usually support a continuous mode. These studies are usually based on point-like atoms, where the wavelength of light is usually much larger than the size of point-like atoms \cite{Wallraff1622004,Roy12017,Goy19031983,Leibfried2812003,Miller5512005,Haroche10832013}. Therefore, we usually use the dipole approximation \cite{Walls2008} to simplify the interaction between photons and atoms when dealing with these systems. With the deepening of quantum optics research and the great technological progress of superconducting circuits  \cite{Manenti9752017,Roy12017,You5892011,Kockum2019,Krantz0213182019}, artificial giant atoms \cite{Roy0210012017,Roy12017,Gustafsson2072014,Kockum2021,Andersson11232019,Kannan7752020} that can interact with surface acoustic waves or microwaves through multiple coupling points have been designed in experiment. Since giant atoms can be designed to couple with waveguides at multiple points with large separation distances, the dipole approximation is no longer valid \cite{Gustafsson2072014}.

In systems containing giant atoms, many interesting and previously undiscovered phenomena resulting from quantum interference effects between multiple coupling points have been predicted, such as the frequency-dependent relaxation rate and the Lamb shift of giant atoms \cite{Kannan7752020,Kockum0138372014,Vadiraj200314167}, decoherence-free interaction between multiple giant atoms \cite{Kannan7752020,Kockum1404042018,Carollo0431842020,Du0237052023,Cilluffo0430702020,Soro023712,Soro0137102023}, non-exponential relaxation \cite{Andersson11232019,Guo053821017,Longhi30172020,Guo0337062020,Du0231982022}, generation of bound states \cite{Guo0430142020,Zhao0538552020,Wang0436022021,Lim0237162023,Xiao802022,Cheng0335222022,Yin0637032022,Vega0535222021}, and electromagnetically induced transparency \cite{Ask201115077,Zhao425062022,Zhu0437102022}, etc. \cite{Du2236022022,Zhang220503674,Noachtar0137022022,Du123012023,Wang0437032022,Brianso0637172022,Santos0536012023,Yang1151042021,Qiu2242122023,Wang0350072022,Burillo0137092020,Wang75802022,Liu742023,Du2236022022,Zhang230316480,Li35982023,Chen230414713,Wang230410710,Du0450102023,Kuo230707949,Jia230402072,Yin230314746,Liu1068542023}. The giant atom scheme provides an effective way to control photons \cite{Wang401162021,Zhang10542992022,Zhang1705682023,Du0237122021,Du30382021,Gu230613836,Cheng3382023,Cai0337102021,Feng0637122021,Zou8968272022,Yin013715}, especially the nonreciprocal propagation of photons \cite{Yu0137202021,Zhou0637032023,Du0537012021,Wang221101819,Du0432262021,Chen2152022,Liu234282022,Sun0351032023}. Moreover, the giant atom scheme can also achieve higher dimensional cold atomic structures in optical lattices \cite{Tudela2036032019}.
In the past, many studies of non-Markovian systems were based on point-like atoms in the traditional framework of quantum optics, where the presence of mirrors or multiple point-like atoms is often required \cite{Eschner4952001,Tufarelli0121132014,Guimond0440122017,Calajo0736012019,Milonni10961974,Zheng1136012013,Ballestero0730152013,Laakso1836012014,Ballestero0423282014,Fang0538452015,Guimond0238082016,Ramos0621042016,Tufarelli0138202013,Dinc2132019}. However, the non-Markovianity of a single giant atom can be achieved by tuning the time delay between adjacent coupling points \cite{Guo053821017,Ask0138402019} in quantum systems containing giant atoms.

The majority of the work on giant atoms deals with two-level or three-level systems coupled to one-dimensional infinite waveguides. In fact, for one-dimensional waveguides, both ends are usually terminated by other media, and light can be partially reflected in waveguides due to the difference in refractive index. The  single-ended and quasi-one-dimensional structures have been realized, for example, by thinning one end of the waveguide to make it almost transparent while the other end is connected to an opaque medium \cite{Claudon1742010,Bleuse1036012011,Reimer7372012,Bulgarini1211062012}. Such a system can be equivalent to an infinite waveguide with a mirror at one end \cite{Barkemeyer19000782020,Song592018,Bradford0638302013,Chang0243052020,Sinha0436032020,Fang043035043035,Zeng0303052023,Barkemeyer0237082022,Crowder0137142022,Xin0537062022,Barkemeyer0337042021,Kabuss102016,Ding230606373,Ding230716876,Wang19002252019,Wang111772019}. However, the non-Markovian dynamics for the two-level giant atom coupled to a semi-infinite waveguide still remains unexplored.

In this paper, we study the non-Markovian dynamics of a two-level giant atom coupled to a one-dimensional semi-infinite waveguide through multiple coupling points, where the finite end of the waveguide is realized by a perfect mirror. When the traveling time of a photon between the adjacent coupling points is sufficiently large compared to the inverse of the bare relaxation rate of the atom, the system enters a non-Markovian process. The photon is continuously transferred between multiple coupling points via the semi-infinite waveguide or is reabsorbed after the photon is reflected by the mirror. We find three different types of bound states, containing the static bound states, the periodic equal and non-equal amplitude oscillating bound states and discuss the physical origins of the bound states formation. We study the atomic dynamics and the corresponding field intensity distributions in the above three cases. Moreover, for a realistic waveguide QED setup in experiment, the light-matter interaction dominates over loss and dephasing, and we discuss the influence of this detrimental phenomenon on the formation of bound states. Finally, we extend the system to a non-Markovian quantum system with many giant atoms and one-dimensional semi-infinite waveguide coupling.

The paper is organized as follows. In Sec.~\ref{er}, we introduce our model, explain methods we used to tackle the non-Markovian dynamics, and give the dynamical equations of the waveguide and giant atom. In Sec.~\ref{san}, we derive the delay differential equation, field intensity function, and exact analytical expression for the probability amplitude of the giant atom. In Sec.~\ref{si}, we discuss the conditions for the formation of bound states and obtain three different types of bound states. Sec.~\ref{wu} gives the dynamical expressions of three different types of bound states and explores the influences of the different parameters on field intensity distribution. Moreover, the bound states under the different number of coupling points are also considered. In Sec.~\ref{liu}, we study the influences of the dissipation of unwanted modes and dephasing on the non-Markovian dynamics of the giant atom. In Sec.~\ref{qi}, we generalize the system to the case containing many two-level giant atoms coupling to a one-dimensional semi-infinite waveguide and derive a set of delay differential equations of the probability amplitude with different giant atoms. In Sec.~\ref{ba}, all of the above work is summarized.

\begin{figure}
\centering\scalebox{0.28}{\includegraphics{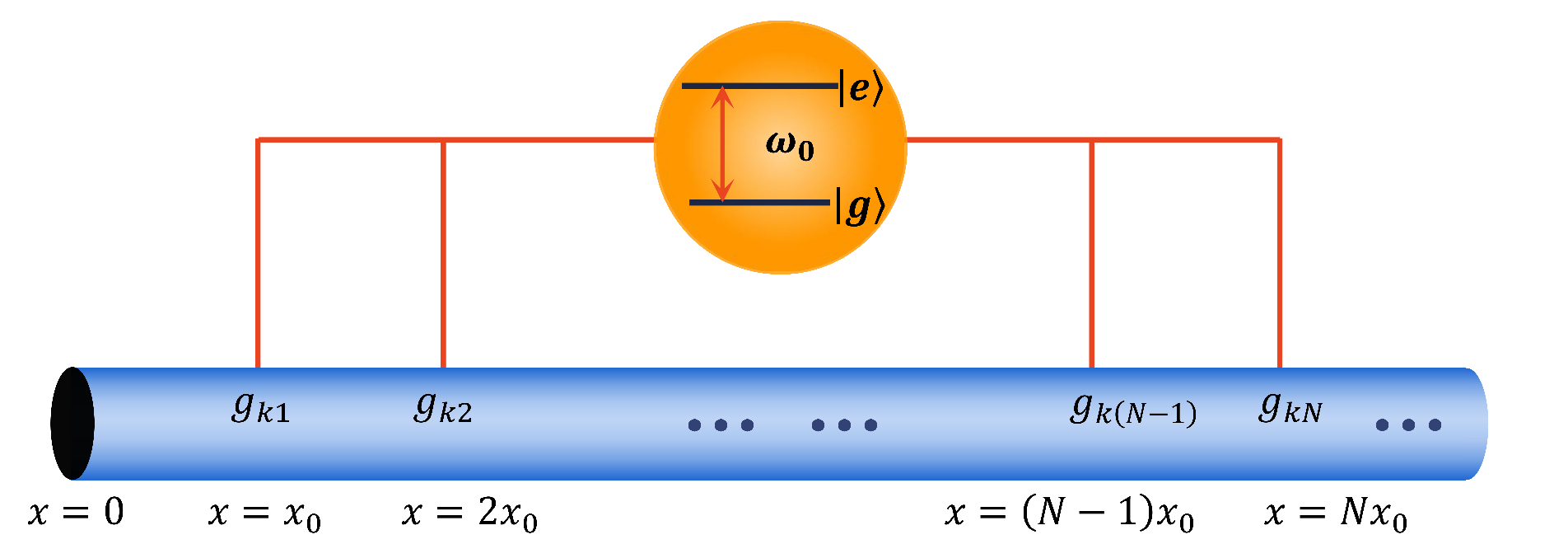}}
\caption{Schematic diagram of the scheme. A two-level giant atom is coupled to a one-dimensional semi-infinite waveguide via $N$ coupling points with coupling coefficient $g_{km}$, where the waveguide is terminated by a perfect mirror with reflectivity $R = 1$ at $x = 0$. The distance between the adjacent coupling points is the same as that between the mirror and the first coupling point, both of which are ${{x_0}}$. $\omega_0$ denotes the transition frequency between the ground state $|g\rangle$ and the excited state $|e\rangle$ of the giant atom.}\label{setup}
\end{figure}

\section{Model Hamiltonian}
\label{er}
We consider a two-level giant atom coupled to a one-dimensional semi-infinite waveguide through $N$ coupling points, which can be realized by photonic crystal waveguide \cite{Wang4201832018,Wang8168702018,Lodahl4306542004} and microwave transmission line \cite{Roy12017,Peropadre0638342011,Mok0538612020,Boch1705012014,Eichler0321062012}. As shown in Fig.~\ref{setup}, one end of the waveguide $(x = 0)$ is terminated by a perfect mirror with the reflectivity $R = 1$, allowing the photon emitted through the atom to transfer between the mirror and each coupling point via
the waveguide. The dispersion relationship of the photon with the wave vector $k$ in the waveguide is approximately linear around the transition frequency ${\omega _0}$ as ${\Omega _k} \simeq {\omega _0} + v(k - {k_0})$ \cite{Roy0210012017,Shen3020012005,Shen0238372009,Shen0238382009,Shen2130012005}, where $v$ is the photon group velocity, and ${\Omega _{{k_0}}} = {\omega _0}$. Any $k>0$ in the infinite waveguide corresponds to two orthogonal static modes with spatial profiles of $ \propto \cos (kx)$ and $ \propto \sin (kx)$, respectively. However, for a semi-infinite waveguide, we only need to consider the sinelike modes because there is a perfect mirror at one end of the waveguide. Therefore, the giant atom interacts with the waveguide at the $m\mathrm{th}$ coupling point with coupling strength ${g_{km}} \propto \sin (km{x_0})(m = 1,~2,~...,~N)$. Under the rotating-wave approximation, the Hamiltonian of the system reads $(\hbar  = 1)$
\begin{equation}
\begin{aligned}
   \hat{H} &= {\hat{H}_a} + {\hat{H}_w} + {\hat{H}_{aw}},\\
   {\hat{H}_a} &= {\omega _0}\left| e \right\rangle \left\langle e \right|,\\
   {\hat{H}_w} &= \int_0^{{k_c}} {dk} {\Omega _k}\hat a_k^\dag {{\hat a}_k},\\
   {\hat{H}_{aw}} &= \sum\limits_{m = 1}^N {\int_0^{{k_c}} {dk} ({g_{km}\hat a_k^\dag{\sigma _ - } }+{g^*_{km}}\hat a_k{\sigma _ + }  )},
\end{aligned}
\label{H}
\end{equation}
where ${\hat{H}_a}$ and ${\hat{H}_w}$ denote the free Hamiltonian of the atom and waveguide, respectively. ${\hat{H}_{aw}}$ describes the interaction between the atom and the semi-infinite waveguide. ${{k_c}}$ stands for a cutoff wave vector depending on a specific waveguide. We define the raising operator (lowering operator) of the atom as ${\sigma _ + } = \left| e \right\rangle \left\langle g \right|({\sigma _ - } = \left| g \right\rangle \left\langle e \right|)$, where $\left| e \right\rangle $ and $\left| g \right\rangle $ denote excited and ground states, respectively. $\hat a_k^\dag ({{\hat a}_k})$ is the photon generation (annihilation) operator, which satisfies the $[ {{{\hat a}_k},\hat a_{k'}^\dag } ] = \delta (k - k')$. Moreover, we set the distance between adjacent coupling points equal to ${x_0}$ (the distance between the endpoint of a semi-infinite waveguide and the first coupling point is also ${x_0}$). Therefore, the time for the photon to travel between any two adjacent coupling points (including between the endpoint of a semi-infinite waveguide and the first coupling point) is a constant ${\tau _0} = {x_0}/v$. In this paper, we explore the non-Markovian dynamics caused by the time delay ${\tau _0}$ induced through adjacent coupling points and the semi-infinite waveguide for the giant atom.

We choose $\left| {e,0} \right\rangle $ as the initial state, where the atom is in the excited state $\left| e \right\rangle $, while the field in the waveguide remains in the vacuum state $\left| 0 \right\rangle $. As we work under the rotating-wave approximation, the processes we are interested in involve a narrow range of wave vectors around $k = {k_0}$, and the boundary of the integral can be changed to $\int_0^{{k_c}} {dk}  \to \int_{ - \infty }^{ + \infty } {dk} $ \cite{Roy0210012017,Shen3020012005,Zhang0323352020,Liao0630042016}. Since the total number of atomic and field excitations in the system is conserved, the state of the total system in the single excitation subspace can be written as
\begin{equation}
   \left| {\psi (t)} \right\rangle  = \varepsilon (t)\left| {e,0} \right\rangle  + \int {dk\varphi (k,t)} \hat a_k^\dag \left| {g,0} \right\rangle,
   \label{psi}
\end{equation}
where $\varepsilon (t)$ denotes the probability amplitude of the giant atom in the excited state $\left| e \right\rangle $, while the second term on the right side of Eq.~(\ref{psi}) describes the state of a single boson propagating in the waveguide with the probability amplitude $\varphi (k,t)$.
Substituting Eqs.~(\ref{H}) and (\ref{psi}) into Schr\"odinger equation $i\partial_t\left|\psi\right\rangle  = \hat{H}\left| \psi  \right\rangle $, we get the set of differential equations of the probability amplitudes
\begin{eqnarray}
   \dot \varepsilon (t) &=&  - i{\omega _0}\varepsilon (t) - i\sum\limits_{m = 1}^N {\int {dk{g_{km}^*}\varphi (k,t)} } ,
   \label{epsdot}\\
   \dot \varphi (k,t) &=&  - i{\Omega _k}\varphi (k,t) - i\sum\limits_{m = 1}^N {g_{km}\varepsilon (t)} ,
   \label{phidot}
\end{eqnarray}
which determine the dynamical evolutions for the giant atom and the semi-infinite waveguide.

\section{Equation of motion and the solution}
\label{san}
In this section, we derive the analytical expressions for the probability amplitude of the two-level giant atom by solving a set of delay differential equations. We set the coupling strength as ${g_{km}} = \sqrt {\Gamma v/\pi } \sin (km{x_0})$~\cite{Tufarelli0121132014}, where $\Gamma $ denotes the spontaneous emission rate of the atom without a mirror. By substituting the coupling strength ${g_{km}}$ and initial condition $\varphi (k,0) = 0$ into Eqs.~(\ref{epsdot}) and (\ref{phidot}),
 we can obtain the atomic excitation probability amplitude
\begin{equation}
\begin{aligned}
   \dot \varepsilon (t) = & - i{\omega_0}\varepsilon (t) \\&- \frac{\Gamma }{2}\sum\limits_{m,n = 1}^N {\varepsilon (t - \left| {m - n} \right|{\tau _0})\Theta (t - \left| {m - n} \right|{\tau _0})} \\ &+ \frac{\Gamma }{2}\sum\limits_{m,n = 1}^N {\varepsilon \left[ {t - (m + n){\tau _0}} \right]\Theta \left[ {t - (m + n){\tau _0}} \right]} ,
   \label{epsdottime}
\end{aligned}
\end{equation}
where $\Theta (x)$ denotes the Heaviside step function, which describes the delayed feedback from the coupling points and the reflection of the semi-infinite waveguide. The first term on the right side of Eq.~(\ref{epsdottime}) describes atomic coherent dynamics. The second term on the right side of Eq.~(\ref{epsdottime}) indicates that the photon transfers from the $m$th coupling point to the $n$th coupling point without being reflected by the mirror, including the Markovian approximation that the photon released from the $m$th coupling point is reabsorbed at the same point without the reflection of the mirror. The last term on the right side of Eq.~(\ref{epsdottime}) denotes the feedback term that the photon released from the $m$th coupling point is absorbed by the $n$th coupling point after being reflected through the mirror. The atomic reabsorption of the emitted photon denoted in the second and third terms of Eq.~(\ref{epsdottime}) occurs at the evolution time $t \geqslant \left| {m - n} \right|{\tau _0}$ and $t \geqslant (m + n){\tau _0}$, respectively. Moreover, we find the probability amplitude with the giant atom in Eq.~(\ref{epsdottime}) can be reduced to that with the point-like atom in Ref.~\cite{Tufarelli0121132014} when $m=n=1$.

In addition to the dynamics of the atom, the dynamical characteristics of the output field are also crucial. The annihilation operator of the photon in real space is $\hat C(x) = \sqrt {2/\pi } \int {dk} {{\hat a}_k}\sin (kx)$, and $\hat C(x)\left| {\psi (t)} \right\rangle  = \phi (x,t)\left| {g,0} \right\rangle $ can be obtained by substituting $\hat C(x)$ into Eq.~(\ref{psi}). Therefore, the probability amplitude $\phi (x,t) = \sqrt {2/\pi } \int {dk} \varphi (k,t)\sin (kx)$ of the real space field is
\begin{equation}
\begin{aligned}
   \phi (x,t) =  &- i\sqrt {\frac{\Gamma }{{2v}}} \sum\limits_{m = 1}^N {\varepsilon ( {t - | {\tau  - m{\tau _0}} |} )\Theta ( {t - | {\tau  - m{\tau _0}} |} )}  \\&+ i\sqrt {\frac{\Gamma }{{2v}}} \sum\limits_{m = 1}^N {\varepsilon [ {t - ( {\tau  + m{\tau _0}} )} ]\Theta [ {t - ( {\tau  + m{\tau _0}} )} ]},
   \label{phitime}
\end{aligned}
\end{equation}
with $\tau = x/v $. The derivation details of Eqs.~(\ref{epsdottime}) and (\ref{phitime}) can be found in Appendix \ref{A}. The field intensity function $P(x,t) = {\left| {\phi (x,t)} \right|^2}$ indicates the probability density at position $x$ and time $t$ to find a single phonon or photon for all possible wave vectors $k$.
Making Laplace transform to Eq.~(\ref{epsdottime}), we get
\begin{equation}
\tilde \varepsilon (s) = \frac{{\varepsilon \left( 0 \right)}}{{s + i{\omega _0} + \frac{\Gamma }{2}\sum\limits_{m,n = 1}^N {\left[ {{e^{ - s\left| {m - n} \right|{\tau _0}}} - {e^{ - s\left( {m + n} \right){\tau _0}}}} \right]} }}.
\label{Eeps}
\end{equation}
\begin{widetext}
Inverting Laplace transformation to Eq.~(\ref{Eeps}) results in
\begin{equation}
\begin{aligned}
 \varepsilon (t) =& \sum\limits_{a,b,c,d,e,f,g,h = 0}^\infty  \mathcal {A}  {e^{ - i{\omega _0} \left[ {t - {\tau _0}(Ne + Na + Nd - f - g + c)} \right]}}
\\&\cdot{\left[ {t - {\tau _0}(Ne + Na + Nd - f - g + c)} \right]^h} \Theta \left[ {t - {\tau _0}(Ne + Na + Nd - f - g + c)} \right],
   \label{epsilon}
\end{aligned}
\end{equation}
with $$\mathcal {A} = {\left(\frac{\Gamma }{2}\right)^h}{( - 1)^{a + b + c + e}}\frac{{2^g}}{{(h - g)!}}\frac{{2h}}{{(2h - e)!e!}}\frac{{(2h + f - 1)!}}{{(g-b)!f!}}\frac{{(g + a - 1)!}}{{(g - 1)!a!}}\frac{{{N^b}}}{{(b - c)!c!}}\frac{{(b + d - 1)!}}{{(b - 1)!d!}},$$
or
\begin{equation}
\varepsilon (t) = \sum\limits_k {\frac{{\varepsilon \left( 0 \right){e^{{s_k}t}}}}{{1 + \frac{\Gamma }{2}\sum\limits_{m,n = 1}^N {\left[ { - \left| {m - n} \right|{\tau _0}{e^{ - {s_k}\left| {m - n} \right|{\tau _0}}} + \left( {m + n} \right){\tau _0}{e^{ - {s_k}\left( {m + n} \right){\tau _0}}}} \right]} }}} ,
   \label{epst}
\end{equation}
\end{widetext}
where Eq.~(\ref{epsilon}) has used the binomial theorem ${(1 + x)^a} = \sum_b^\infty  {\frac{{a!}}{{(a - b)!b!}}} {x^b}$ and ${(1 + x)^{ - a}} = \sum_b^\infty  {\frac{{{{( - 1)}^b}(a + b - 1)!}}{{(a - 1)!b!}}} {x^b}$.
The complex frequency parameters ${s_k}$ in Eq.~(\ref{epst}) are determined by
\begin{equation}
   {s_k} + i{\omega _0} + \frac{\Gamma }{2}\sum\limits_{m,n = 1}^N {\left[ {{e^{ - {s_k}\left| {m - n} \right|{\tau _0}}} - {e^{ - {s_k}\left( {m + n} \right){\tau _0}}}} \right]}  = 0.
\label{sk}
\end{equation}
For finite time delay $\tau_0 > 0$, Eq.~(\ref{sk}) has multiple solutions. We will further investigate the specific form of ${s_k}$ of Eq.~(\ref{sk}) in Sec.~\ref{si}.

\section{Discussion of bound state conditions}
\label{si}
\begin{figure}[t]
\centering\scalebox{0.333}{\includegraphics{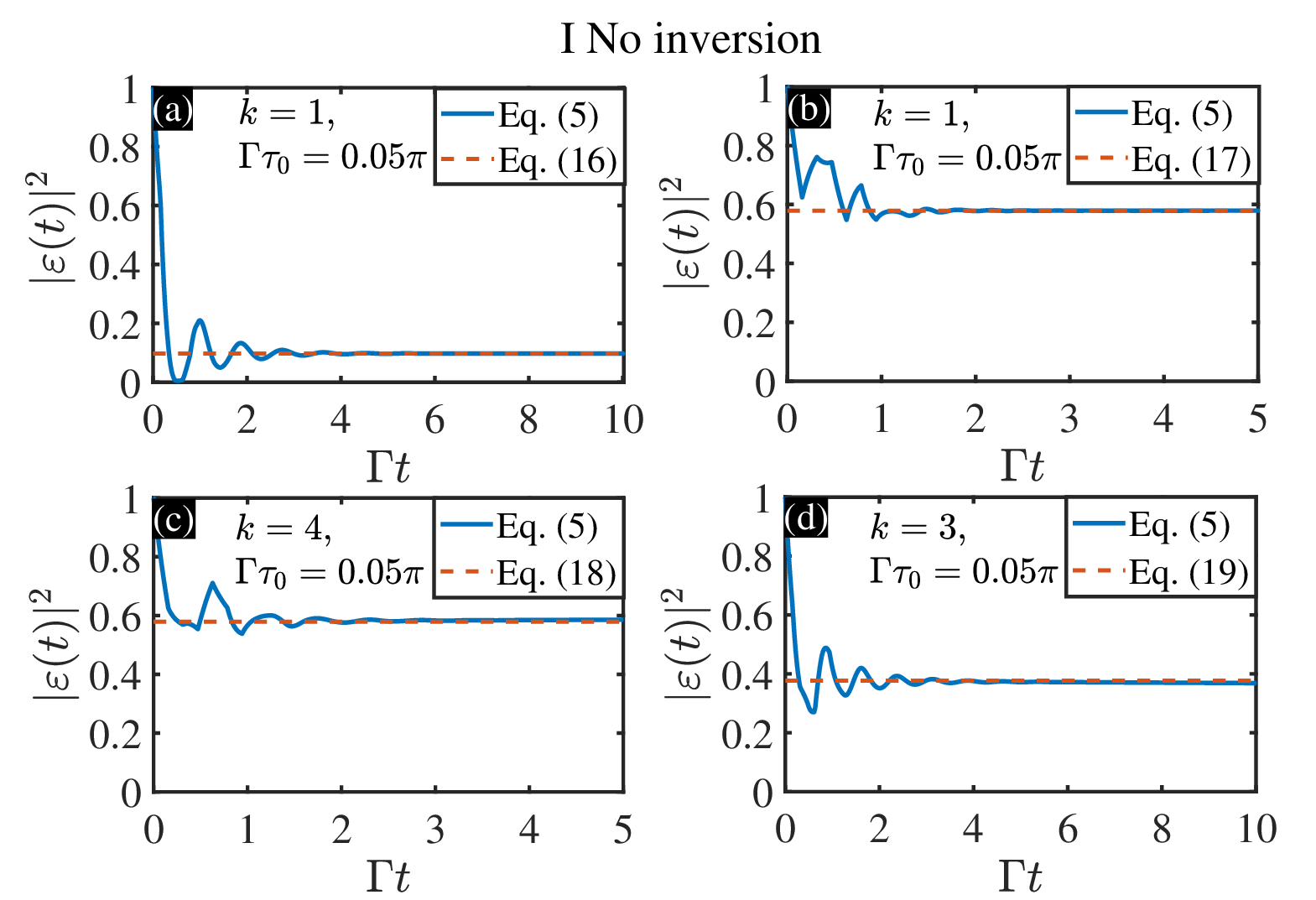}}
\caption{Static bound states for the giant atom with the number of coupling points $N = 3$, the blue-solid and orange-dashed lines correspond to the numerical simulation in  Eq.~(\ref{epsdottime}) and the analytical solutions with Eqs.~(\ref{oneepsk})-(\ref{oneepsN1}), respectively. The parameters chosen are (a) $k = 1, \omega_0\tau_0=2\pi, \Gamma\tau_0 = 0.05\pi$; (b) $k = 1, \omega_0\tau_0=3\pi, \Gamma\tau_0 = 0.05\pi$; (c) $k = 4, \omega_0\tau_0=2.6234\pi, \Gamma\tau_0 = 0.05\pi$; (d) $k = 3, \omega_0\tau_0=1.6\pi, \Gamma\tau_0 = 0.05\pi$.}
\label{one}
\end{figure}
\begin{figure}[t]
\centering\scalebox{0.43}{\includegraphics{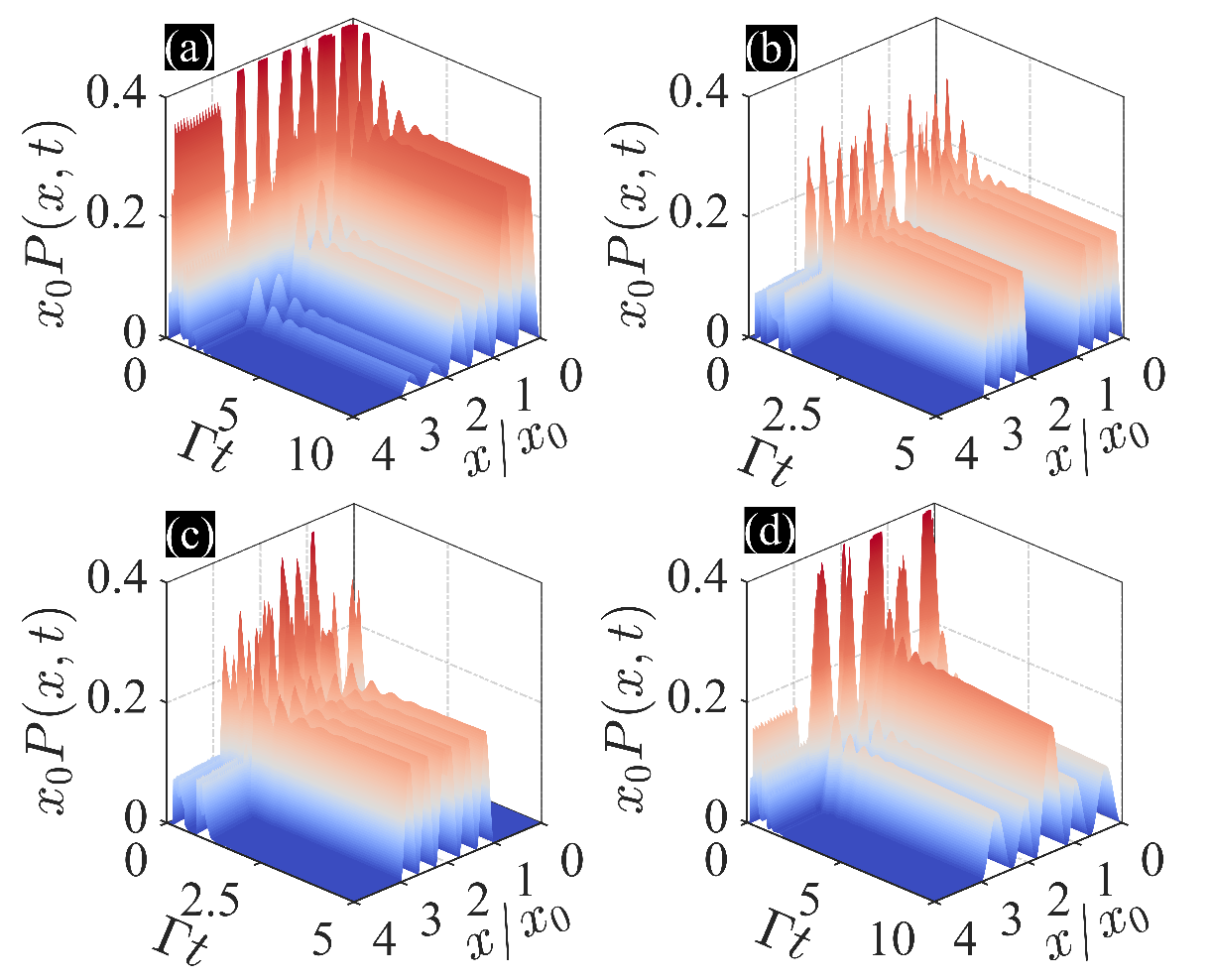}}
\caption{The figure shows the evolution of the field intensity $P(x,t) = {\left| {\phi (x,t)} \right|^2}$ with time and position by solving Eq.~(\ref{phitime}) in the waveguide under four different conditions. The colors indicate the field intensity of the bound states. The parameters chosen are the same as those in Fig.~\ref{one}. }
\label{oneP3}
\end{figure}
In this section, we give the conditions for the generation of bound states and discuss whether these conditions can coexist. $s_k$ in Eq.~(\ref{sk}) is pure imaginary when the real part of the complex frequency $s_k$ denoting the relaxation rate equals $0$. In this case, the corresponding mode is a bound state not decaying despite the dissipative environment. We seek the pure imaginary solution ${s_k} =  - i{\omega _k}$ in Eq.~(\ref{sk}), which can be satisfied when
\begin{subequations}
\label{sk12}
\begin{align}
{s_k} &=  - i{\omega _k} =  - i{\omega _0} =  - i2k\pi /{\tau _0},
\label{sk1}\\
{s_k} &=  - i{\omega _k} =  - i{\omega _0} =  - i(2k + 1)\pi /{\tau _0},
\label{sk2}
\end{align}
\end{subequations}
and the corresponding bound state conditions are
\begin{subequations}
\label{omegak12}
\begin{align}
    {\omega _0}{\tau _0} &= {\omega _k}{\tau _0} = 2k\pi,
\label{omegak1}
    \\{\omega _0}{\tau _0} &= {\omega _k}{\tau _0} = (2k + 1)\pi .
\label{omegak2}
\end{align}
\end{subequations}
In order to find more ${s_k}$ meeting Eq.~(\ref{sk}), we divide Eq.~(\ref{sk}) into real and imaginary parts respectively described through
\begin{small}
\begin{equation}
\begin{aligned}
   &\Gamma {\tau _0}{\csc^2} \left(\frac{{{\omega _k}{\tau _0}}}{2}\right){\sin ^2}\left(\frac{{N{\omega _k}{\tau _0}}}{2}\right){\sin ^2}\left[\frac{{\left(N + 1\right){\omega _k}{\tau _0}}}{2}\right] = 0 \hfill ,
  \\&\Gamma {\tau _0}\sin ({\omega _k}{\tau _0}) + 2N\Gamma {\tau _0}\sin ({\omega _k}{\tau _0}) - 2\Gamma {\tau _0}\sin (N{\omega _k}{\tau _0}) \\&-2\Gamma {\tau _0}\sin \left[ {(N + 1){\omega _k}{\tau _0}} \right] + \Gamma {\tau _0}\sin \left[ {(2N + 1){\omega _k}{\tau _0}} \right] \\&-4({\omega _k}{\tau _0} - {\omega _0}{\tau _0})\left[ {1 - \cos ({\omega _k}{\tau _0})} \right] = 0,
   \label{real}
\end{aligned}
\end{equation}
\end{small}
which can be derived in Appendix \ref{B}.
Solving the two equations in Eq.~(\ref{real}) simultaneously, we obtain
\begin{subequations}
\label{sk34}
\begin{align}
  {s_k} &=  - i{\omega _k} =  - i2k\pi /\left( {N{\tau _0}} \right),
  \label{sk3}\\
  {s_k} &=  - i{\omega _k} =  - i2k\pi /\left[ {\left( {N + 1} \right){\tau _0}} \right],
 \label{sk4}
\end{align}
\end{subequations}
and the corresponding bound state conditions satisfy
\begin{subequations}
\label{omegak34}
\begin{align}
  {\omega _0}{\tau _0} &= \frac{{2k\pi }}{N} - \frac{1}{2}N\Gamma {\tau _0}\cot \left(\frac{{k\pi }}{N}\right),
\label{omegak3}
  \\{\omega _0}{\tau _0} &= \frac{{2k\pi }}{{N + 1}} - \frac{1}{2}\left(N + 1\right)\Gamma {\tau _0}\cot \left(\frac{{k\pi }}{{N + 1}}\right).
\label{omegak4}
\end{align}
\end{subequations}
Under the Markovian limit ($\Gamma {\tau _0} \to 0$), the bound state conditions (\ref{omegak3}) and (\ref{omegak4}) can be reduced to ${\omega _0}{\tau _0} = 2k\pi/N$ and ${\omega _0}{\tau _0} = 2k\pi/(N + 1)$. However, in the non-Markovian regime with the large enough $\Gamma {\tau _0}$, the influence of the cotangent terms in Eqs.~(\ref{omegak3}) and (\ref{omegak4}) can not be ignored. Solving the transcendental equation in Eq.~(\ref{omegak3}) or (\ref{omegak4}) leads to two cases containing only one integer $k$ satisfying it or both two integers $k_1$ and $k_2$ meeting it, whose derivation can be found in Appendix \ref{C}.
To ensure the validity of the rotating-wave approximation, we need to guarantee $\left| {{\omega _k} - {\omega _0}} \right|/{\omega _0} \ll 1$, which is equivalent to $|N\Gamma\cot(k\pi/N)/(2\omega_0)| \ll 1$ and $|(N+1)\Gamma\cot[k\pi/(N+1)]/(2\omega_0)| \ll 1$ with $k \in {\mathbb{Z}^ + }$ according to Eqs.~(\ref{omegak3}) and (\ref{omegak4}).

Based on the number of bound states in the waveguide, we obtain three cases and summarize them as follows:

(i) One of the bound state conditions (\ref{omegak1}), (\ref{omegak2}), (\ref{omegak3}), and (\ref{omegak4}) is satisfied, and there is only one integer $k_0$ as its solution. In this case, there will be four situations given by Eqs.~(\ref{oneepsk})-(\ref{oneepsN1}) with Eqs.~(\ref{sk12}) and (\ref{sk34}) (see Sec.~\ref{wu} for more details).

(ii) Two integer solutions $k_1$ and $k_2$ meet the bound state condition (\ref{omegak3}) or (\ref{omegak4}), which will induce two cases obtained by  Eqs.~(\ref{twoepsN}) and (\ref{twoepsN1}) with Eq.~(\ref{sk34}) in Sec.~\ref{wu}.

(iii) Eq.~(\ref{omegak1}) or (\ref{omegak2}) is satisfied on the basis of meeting the case (ii), which leads to the existence of three integers $k_0$, $k_1$, and $k_2$ simultaneously. The corresponding four results determined by Eqs.~(\ref{threeepsN}), (\ref{threeepsN1}), (\ref{D1}), and (\ref{D2}) with Eqs.~(\ref{sk12}) and (\ref{sk34}) will be studied in Sec.~\ref{wu} and Appendix \ref{C}.

Next, we discuss the dynamics and output field of the giant atom in detail.

\begin{figure}
\centering\scalebox{0.333}{\includegraphics{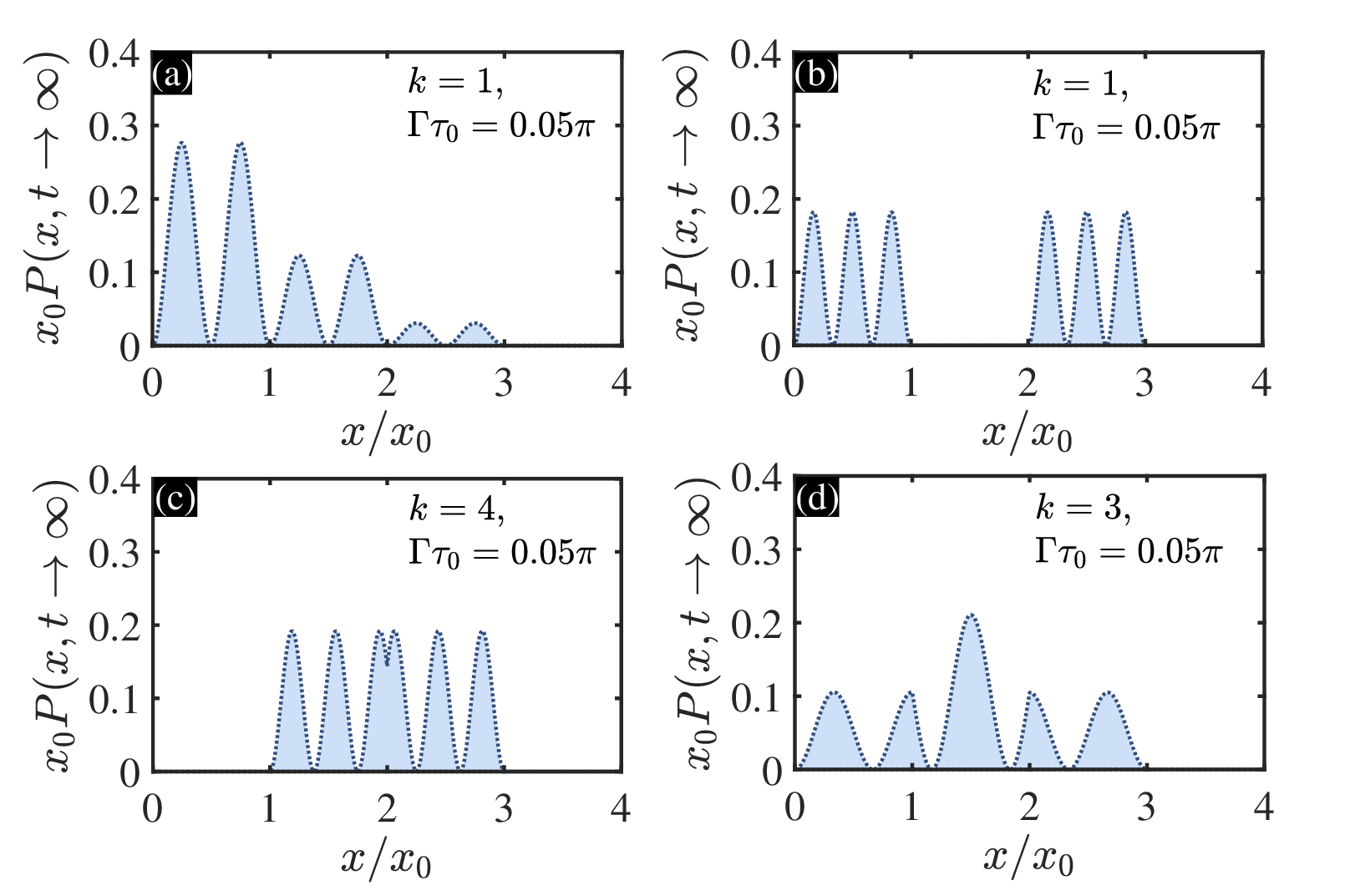}}
\caption{The field intensity distribution $P(x,t)$ in the waveguide at $t\rightarrow\infty$. The four cases shown by (a)-(d) correspond to Figs.~\ref{oneP3}(a)-\ref{oneP3}(d), where the blue-dotted line corresponds to the numerical simulation based on Eq.~(\ref{phitime}). The parameters chosen are the same as those in Fig.~\ref{oneP3}. }
\label{oneP2}
\end{figure}
\begin{figure}
\centering\scalebox{0.335}{\includegraphics{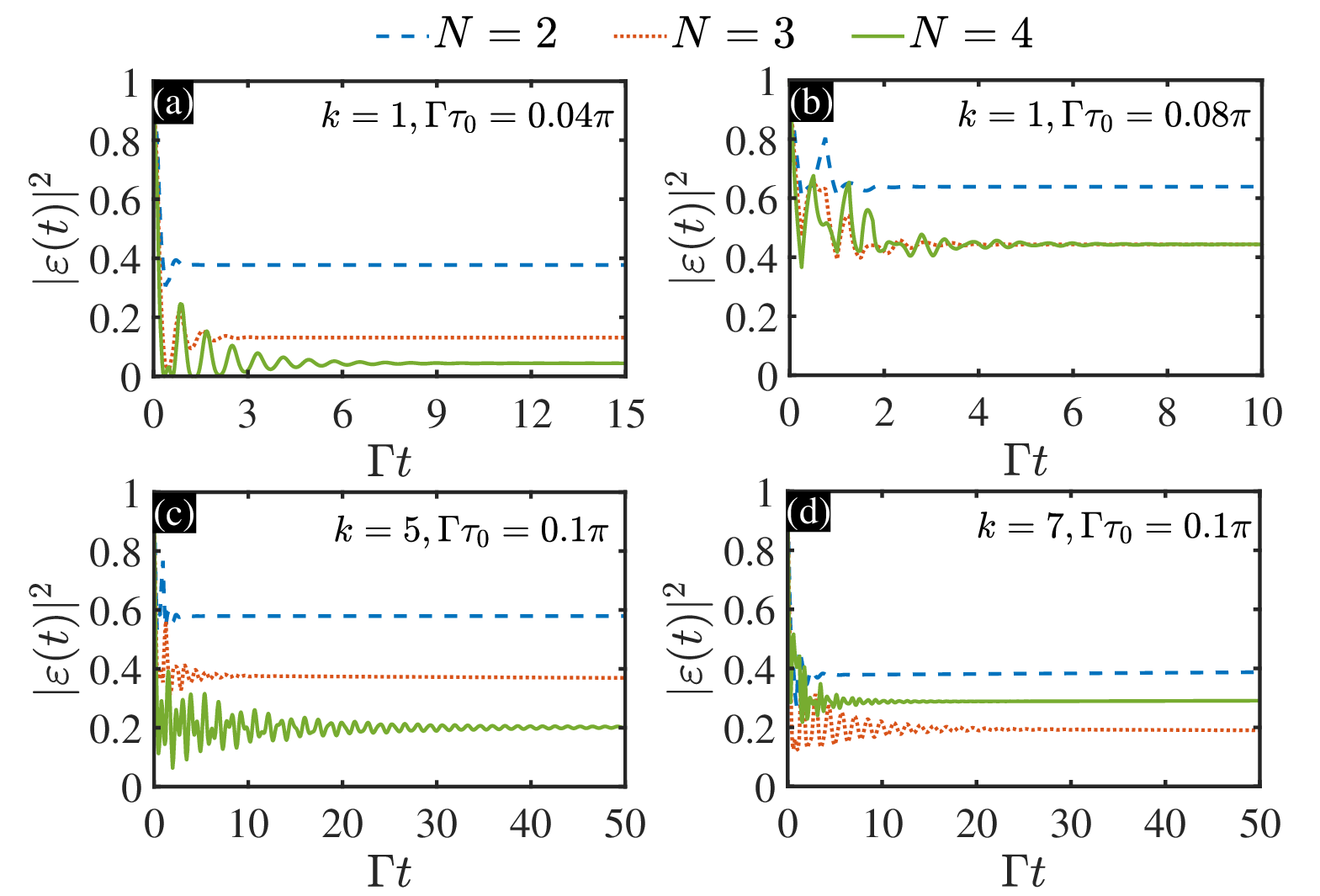}}
\caption{Probability amplitudes in Eq.~(\ref{epsdottime}) as function of time $t$ (in units of $\Gamma^{-1}$) when the number of coupling points $N$ is different. The blue-dashed, orange-dotted, and green-solid lines correspond to $N = 2, N = 3, N = 4$, respectively. The other parameters chosen are (a) $k = 1, \Gamma\tau_0 = 0.04\pi$; (b) $k = 1, \Gamma\tau_0 = 0.08\pi$; (c) $k = 5, \Gamma\tau_0 = 0.1\pi$; (d) $k = 7, \Gamma\tau_0 = 0.1\pi$.}
\label{oneN}
\end{figure}

\section{Bound states in the semi-infinite waveguide}
\label{wu}
We study the non-Markovian dynamics in the three cases obtained from the discussion of the bound state conditions in Sec.~\ref{si}. For the sake of clearness, we divide them into the static bound states with the no inversion of population, the periodic equal amplitude oscillation with two bound states, and the periodic non-equal amplitude oscillation with three bound states. In the following sections, we will discuss them and give the physical origins of the bound states formation.

\begin{widetext}
\subsection{No inversion of population}

We consider the static bound state satisfying case (i) in Sec.~\ref{si}, which means that the bound state in the waveguide has only one frequency. Substituting four $s_k$ in Eqs.~(\ref{sk12}) and (\ref{sk34}) into Eq.~(\ref{epst}), the long-time dynamics of the atomic excitation probability amplitude for four cases reads
\begin{eqnarray}
\varepsilon (t) &=& \frac{1}{{1 + \frac{\Gamma }{2}\sum\limits_{m,n = 1}^N {\left[ { - \left| {m - n} \right|{\tau _0} + \left( {m + n} \right){\tau _0}} \right]} }}{e^{ - i2k\pi t/{\tau _0}}},
 \label{oneepsk}\\
\varepsilon (t) &=& \frac{1}{{1 + \frac{\Gamma }{2}\sum\limits_{m,n = 1}^N {\left[ { - \left| {m - n} \right|{\tau _0}{{\left( { - 1} \right)}^{\left| {m - n} \right|}} + \left( {m + n} \right){\tau _0}{{\left( { - 1} \right)}^{\left( {m + n} \right)}}} \right]} }}{e^{ - i(2k + 1)\pi t/{\tau _0}}},
   \label{oneepsk1}\\
\varepsilon (t) &=& \frac{{2{{\sin }^2}({\omega _k}{\tau _0}/2)}}{{2{{\sin }^2}({\omega _k}{\tau _0}/2) + N\Gamma {\tau _0}}}{e^{ - i2k\pi t/\left( {N{\tau _0}} \right)}},
   \label{oneepsN}\\
\varepsilon (t) &=& \frac{{2{{\sin }^2}({{\tilde \omega }_k}{\tau _0}/2)}}{{2{{\sin }^2}({{\tilde \omega }_k}{\tau _0}/2) + (N + 1)\Gamma {\tau _0}}}{e^{ - i2k\pi t/\left[ {(N + 1){\tau _0}} \right]}},
   \label{oneepsN1}
\end{eqnarray}
\end{widetext}
where ${\omega _k} = 2k\pi /\left( {N{\tau _0}} \right)$ and ${{\tilde \omega }_k} = 2k\pi /\left[ {(N + 1){\tau _0}} \right]$.
\begin{figure}[t]
\centering\scalebox{0.331}{\includegraphics{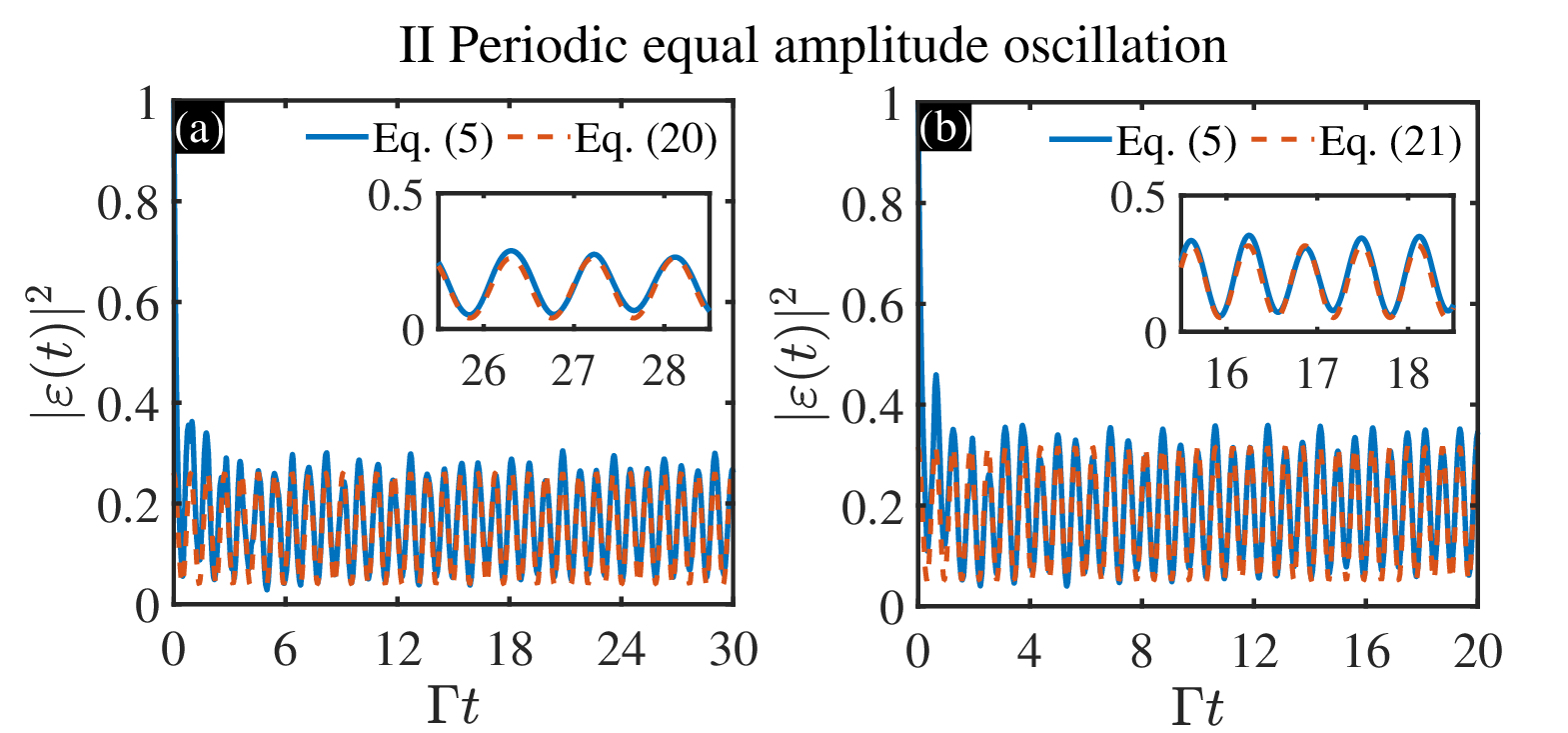}}
\caption{Periodic equal amplitude oscillating bound states in the waveguide for a giant atom with the number of coupling points $N=6$. The orange-dashed and blue-solid lines correspond to the analytical solutions in Eqs.~(\ref{twoepsN}) and (\ref{twoepsN1}) and numerical simulation of Eq.~(\ref{epsdottime}), respectively. The parameters chosen are (a) $k_{1} = 23, k_{2} = 26, \omega_0\tau_0=8.4167\pi, \Gamma\tau_0=0.1443\pi$; (b) $k_{1} = 27, k_{2} = 30, \omega_0\tau_0=8.3336\pi, \Gamma\tau_0=0.0852\pi$. $\omega_{k_n}$ in (a) and (b) are determined by Eqs.~(\ref{sk3}) and (\ref{sk4}), respectively. The insets in the figure magnify the probability amplitudes.}
\label{two}
\end{figure}
\begin{figure}[t]
\centering\scalebox{0.334}{\includegraphics{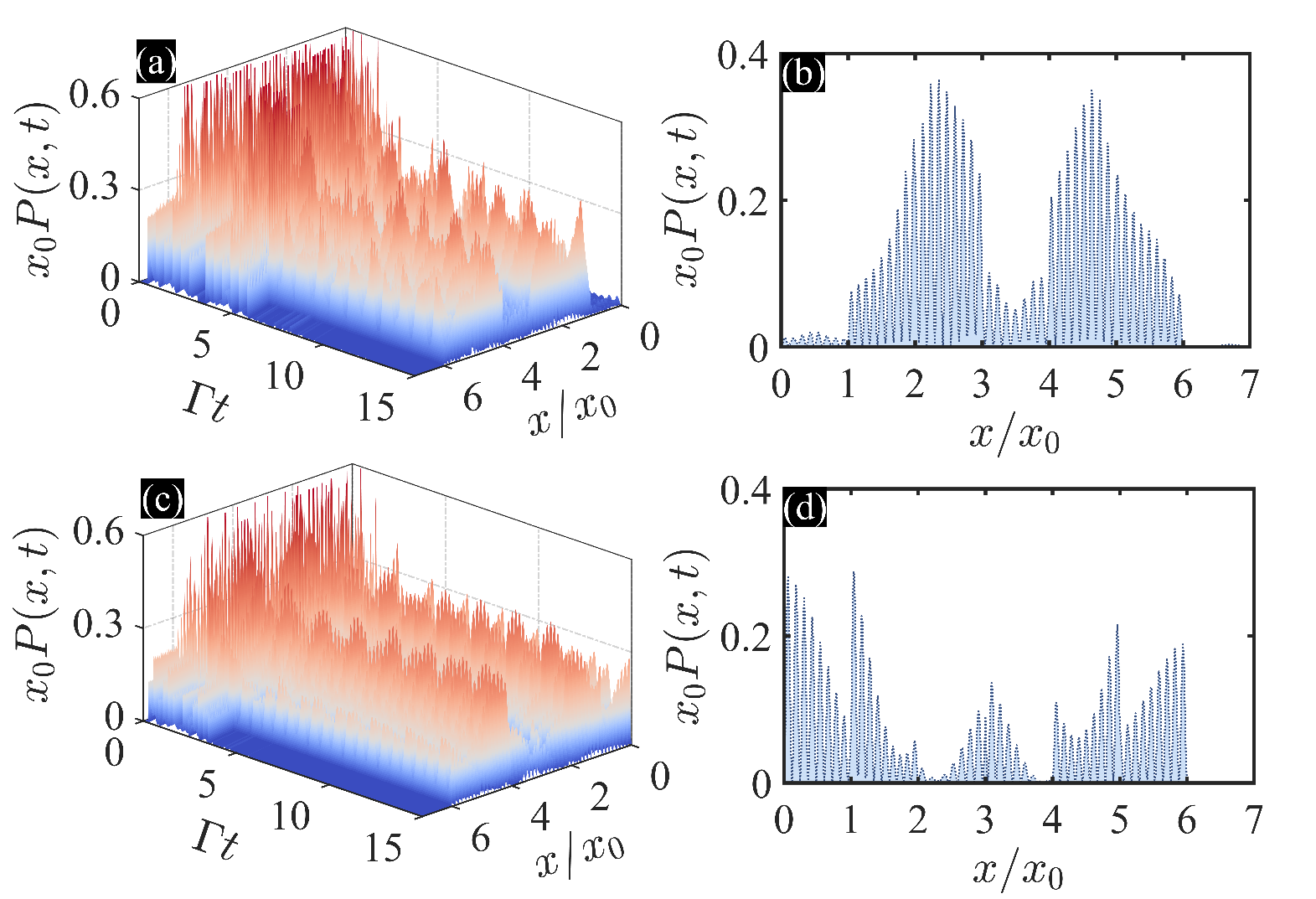}}
\caption{Plot of the field intensity function $P(x,t)$ solved by Eq.~(\ref{phitime}) as a function of $\Gamma t$ and $x/x_0$. (a) and (c) show the evolution of the field intensity in the waveguide corresponding to Fig.~\ref{two}. The colors in (a) and (c) show the field intensity of the bound states. (b) and (d) denote the field intensity distribution $P(x,t)$ in the waveguide at fixed parameter $t=15/\Gamma$ corresponding to (a) and (c), respectively, where the blue-dotted line is the numerical simulation based on Eq.~(\ref{phitime}). The other parameters are the same as those in Fig.~\ref{two}.}
\label{twoP32}
\end{figure}

To compare the analytical solutions given by Eqs.~(\ref{oneepsk})-(\ref{oneepsN1}) with the numerical simulation of Eq.~(\ref{epsdottime}), we plot the atomic excitation probability with $N=3$ as a function of time $t$ (in units of $\Gamma^{-1}$) in Fig.~\ref{one}. The blue-solid and orange-dashed lines indicate the numerical and analytical results, respectively. We find that the analytical expressions show good agreement with those obtained by the numerical simulations with different parameters. In Fig.~\ref{one}, the atomic excitation probabilities ${\left| {\varepsilon (t)} \right|^2}$ in the four cases finally hold a nonzero steady value after a long time, which means that the photon is captured and forms a bound state. This originates
from the transferring of the photon in multiple coupling points and reflecting via the semi-infinite waveguide.

To observe how the bound state is formed, we take the field intensity function $P(x,t)$ as a function of the time $t$ (in units of $\Gamma^{-1}$) and the position $x$ (in units of $x_0^{-1}$). The time evolution of the field intensity function for the four different bound states with the number of coupling points $N=3$ is shown in Fig.~\ref{oneP3}. We observe that the field intensity outside the last coupling point ($x > 3x_0$) disappears with time, while the field intensity at $x < 3x_0$ forms a steady state, where a static bound state has been formed in the waveguide. In Fig.~\ref{oneP2}, we plot the long-time field intensity distribution corresponding to Fig.~\ref{oneP3}, where the blue-dashed line denotes the numerical simulation given by Eq.~(\ref{phitime}). In Fig.~\ref{oneP2}, the field intensities in the above four cases are distributed between the last coupling point and the mirror at the endpoint of the semi-infinite waveguide with $\Gamma t \to \infty $.

The variation of the atomic excitation probability $|\varepsilon(t)|^2$ with respect to the $\Gamma t$ for different number coupling points $N$ is plotted in Fig.~\ref{oneN}. We find that the atomic excitation probability is a steady value although it changes with the number of coupling points. This means that the static bound states with different probabilities can be got through tuning the number of coupling points $N$.

\subsection{Periodic equal amplitude oscillating bound states}

When case (ii) in Sec.~\ref{si} is met, two bound states with frequencies $\omega_{k_1}$ and $\omega_{k_2}$ exist simultaneously in the system. Substituting ${s_k} = \{  - i2k\pi /\left( {N{\tau _0}} \right), - i2k\pi /\left[ {(N + 1){\tau _0}} \right]\}$ obtained by Eq.~(\ref{sk34}) into Eq.~(\ref{epst}), the long-time atomic excitation probability amplitudes are given by
\begin{equation}
\begin{aligned}
\varepsilon (t) =& \frac{{2{{\sin }^2}({\omega _{{k_1}}}{\tau _0}/2)}}{{2{{\sin }^2}({\omega _{{k_1}}}{\tau _0}/2) + N\Gamma {\tau _0}}}{e^{ - i2{k_1}\pi t/\left( {N{\tau _0}} \right)}}\\ &+ \frac{{2{{\sin }^2}({\omega _{{k_2}}}{\tau _0}/2)}}{{2{{\sin }^2}({\omega _{{k_2}}}{\tau _0}/2) + N\Gamma {\tau _0}}}{e^{ - i2{k_2}\pi t/\left( {N{\tau _0}} \right)}},
   \label{twoepsN}
\end{aligned}
\end{equation}
where ${\omega _{{k_n}}} = 2{k_n}\pi /\left( {N{\tau _0}} \right)~(n = 1,~2)$, and
\begin{equation}
\begin{aligned}
\varepsilon (t) =& \frac{{2{{\sin }^2}({\omega _{{k_1}}}{\tau _0}/2)}}{{2{{\sin }^2}({\omega _{{k_1}}}{\tau _0}/2) + (N + 1)\Gamma {\tau _0}}}{e^{ - i2{k_1}\pi t/\left[ {(N + 1){\tau _0}} \right]}} \\&+ \frac{{2{{\sin }^2}({\omega _{{k_2}}}{\tau _0}/2)}}{{2{{\sin }^2}({\omega _{{k_2}}}{\tau _0}/2) + (N + 1)\Gamma {\tau _0}}}{e^{ - i2{k_2}\pi t/\left[ {(N + 1){\tau _0}} \right]}},
   \label{twoepsN1}
\end{aligned}
\end{equation}
with ${\omega _{{k_n}}} = 2{k_n}\pi /\left[ {(N + 1){\tau _0}} \right]~(n = 1,~2)$.
\begin{figure}[t]
\centering\scalebox{0.332}{\includegraphics{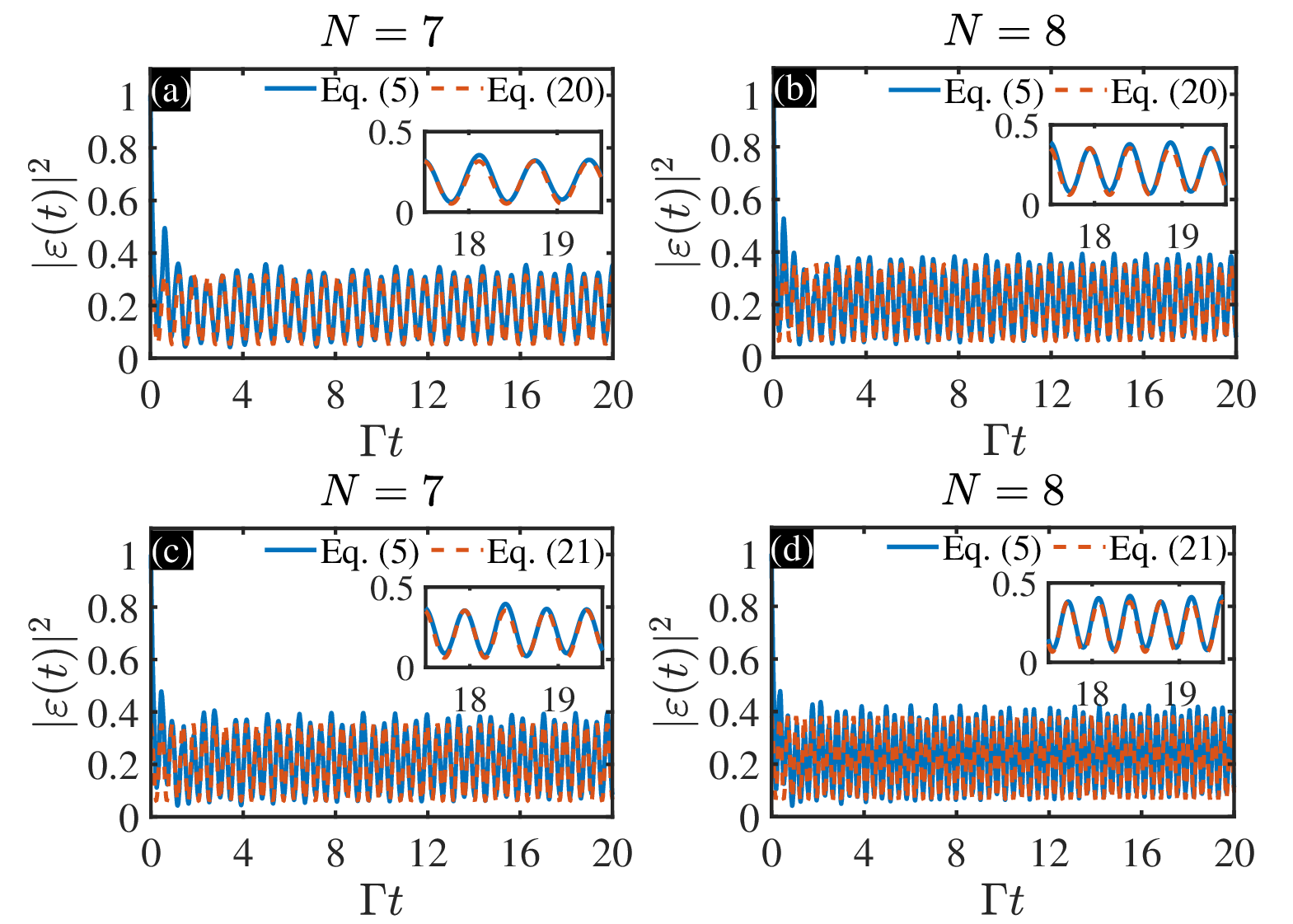}}
\caption{The influence of different number of coupling points $N$ on the equal amplitude oscillating bound states for the giant atom. The orange-dashed and blue-solid lines correspond to the analytical solutions in Eqs.~(\ref{twoepsN}) and (\ref{twoepsN1}) and numerical simulations based on Eq.~(\ref{epsdottime}), respectively. The parameters chosen are (a) $N=7, k_{1} = 27, k_{2} = 30, \omega_0\tau_0=8.3336\pi, \Gamma\tau_0=0.0852\pi$; (b) $N=8, k_{1} = 31, k_{2} = 34, \omega_0\tau_0=8.2803\pi, \Gamma\tau_0=0.0549\pi$; (c) $N=7, k_{1} = 31, k_{2} = 34, \omega_0\tau_0=8.2803\pi, \Gamma\tau_0=0.0549\pi$; (d) $N=8, k_{1} = 35, k_{2} = 38, \omega_0\tau_0=8.2428\pi, \Gamma\tau_0=0.0376\pi$. $\omega_{k_n}$ in (a)-(b) and (c)-(d) are determined by Eqs.~(\ref{sk3}) and (\ref{sk4}), respectively.}
\label{twoN}
\end{figure}
\begin{figure}
\centering\scalebox{0.323}{\includegraphics{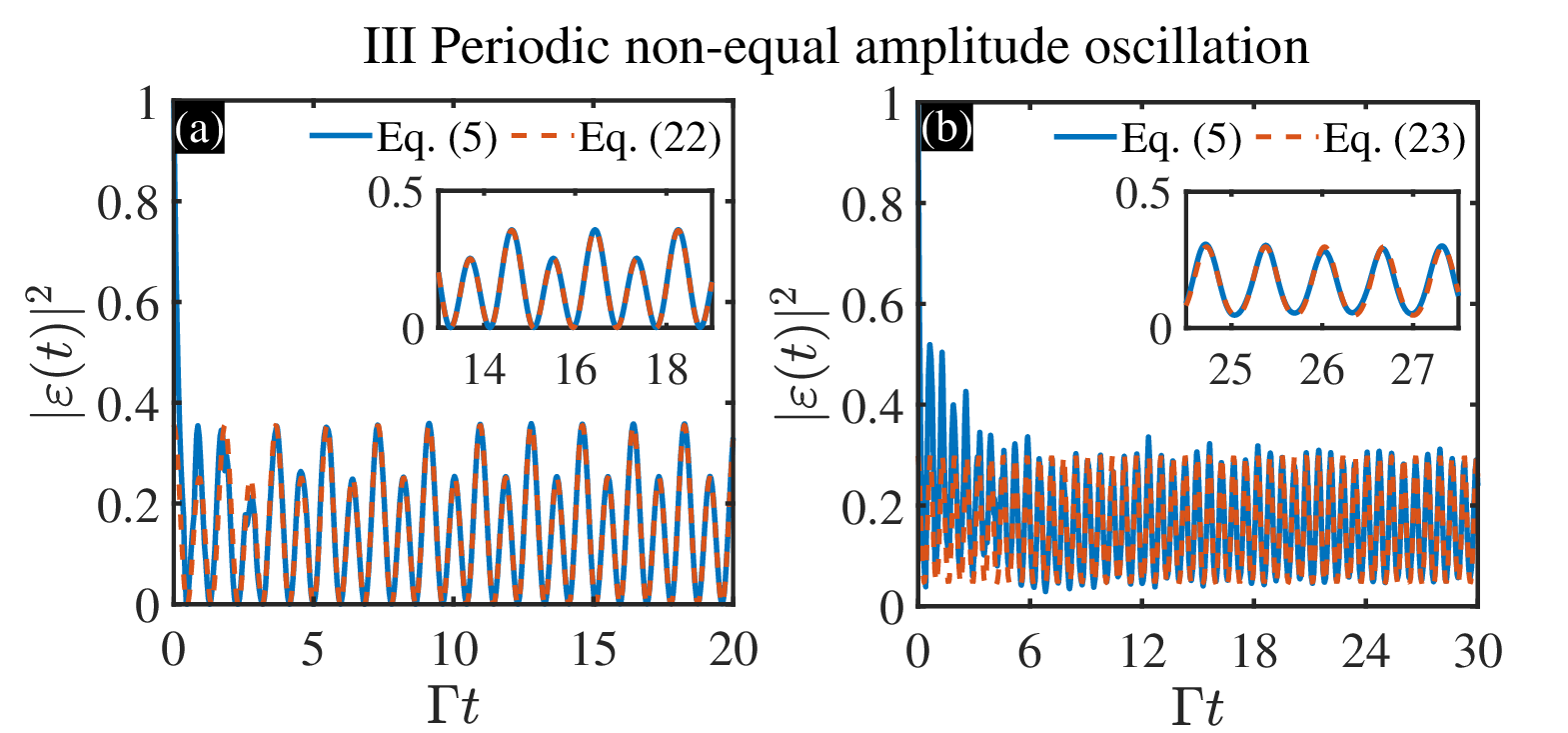}}
\caption{Non-equal amplitude oscillating bound states of a giant atom. The orange-dashed and blue-solid lines correspond to the analytical solutions in Eqs.~(\ref{threeepsN}) and (\ref{threeepsN1}) and numerical simulation solved by Eq.~(\ref{epsdottime}), respectively. The parameters chosen are (a) $N=5, k_{0}=4, k_{1} = 19, k_{2} = 21, \omega_0\tau_0=8\pi, \Gamma\tau_0=0.1162\pi$; (b) $N=8, k_{0}=8, k_{1} = 63, k_{2} = 73, \omega_0\tau_0=17\pi, \Gamma\tau_0=0.1294\pi$. $\omega_{k_n}$ in (a) and (b) are determined by Eq.~(\ref{sk3}).}
\label{three}
\end{figure}
\begin{figure}
\centering\scalebox{0.335}{\includegraphics{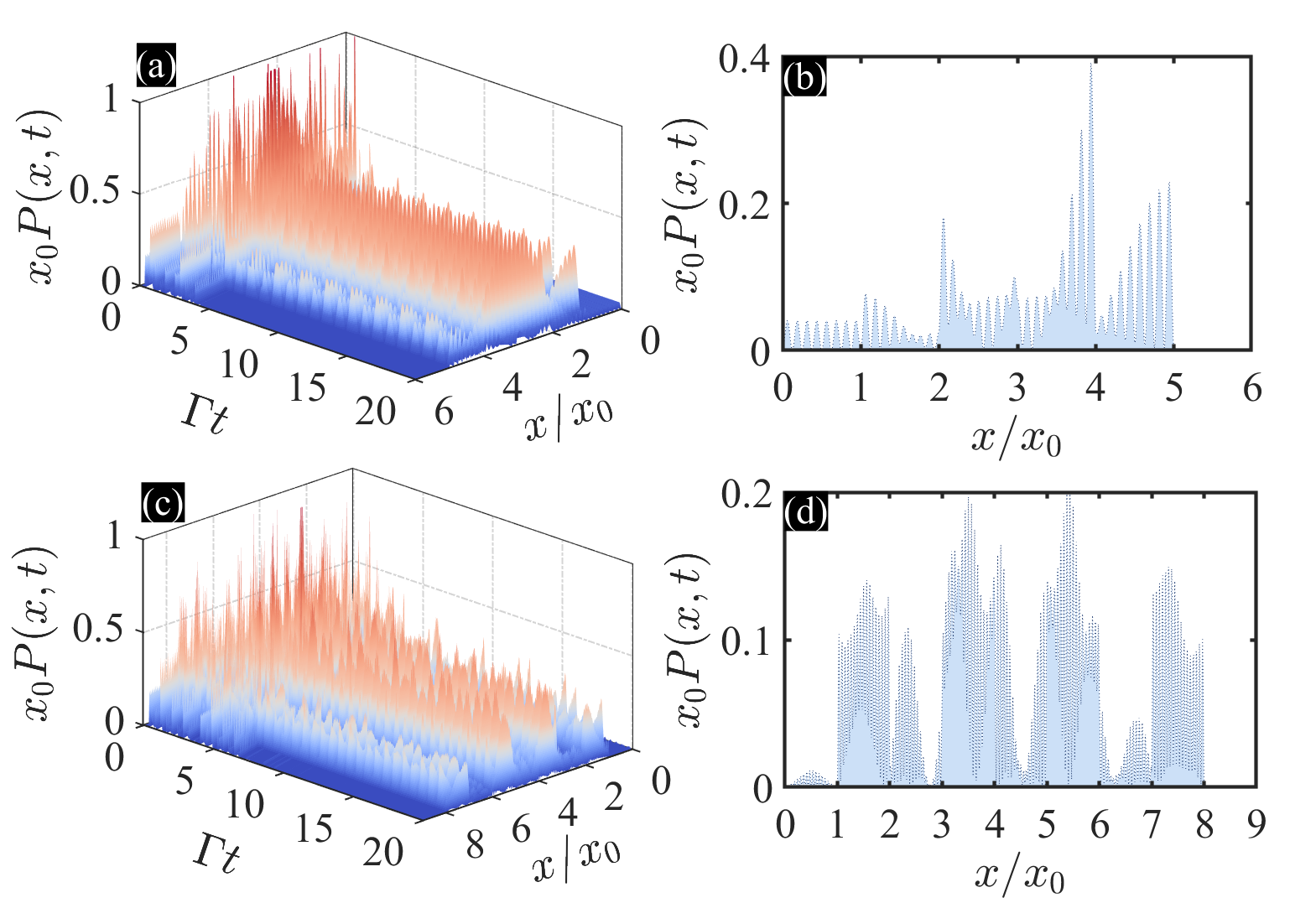}}
\caption{The field strength $P(x,t) = {\left| {\phi (x,t)} \right|^2}$ as a function of $\Gamma t$ and $x/x_0$ based on Eq.~(\ref{phitime}). (a) and (c) show the time evolution of the field intensity $P(x,t)$ in the waveguide corresponding to Figs.~\ref{three}(a) and \ref{three}(b), where the parameters $k_0, k_1,$ and $k_2$ are the same as those in Fig.~\ref{three}. The colors in (a) and (c) describe the field intensity of the bound states. For clarity, (b) and (d) show the distribution of $P(x,t)$ in the waveguide at $t=20/\Gamma$, which respectively correspond to (a) and (c). The blue-dotted line corresponds to the numerical simulation of Eq.~(\ref{phitime}).}
\label{threeP32}
\end{figure}
In this case, two integers $k_1$ and $k_2$ that satisfy transcendental equation (\ref{omegak3}) or (\ref{omegak4}) can be found. The atomic excitation probability ${\left| {\varepsilon (t)} \right|^2}$ is the superposition of the bound states with different frequencies ${\omega _{{k_1}}}$ and ${\omega _{{k_2}}}$. In Appendix \ref{C}, we give $\omega_0 \tau_0$ and $\Gamma \tau_0$ for the existence of two bound states. The long-time dynamics in Eq.~(\ref{twoepsN}) is consistent with that given in Refs.~\cite{Guo0430142020,Lim0237162023}. However, in addition to Eq.~(\ref{twoepsN}), another solution in Eq.~(\ref{twoepsN1}) can be sought out, which is significantly different from that in Refs.~\cite{Guo0430142020,Lim0237162023} and induced by the reflection of the semi-infinite waveguide and time delay between multiple coupling points. This indicates that the atomic excitation probability can be manipulated due to the existence of the mirror in the semi-infinite waveguide without changing the number of coupling points.

We plot the numerical result based on Eq.~(\ref{epsdottime}) and analytical results in  Eqs.~(\ref{twoepsN}) and (\ref{twoepsN1}) for the atomic excitation probability denoted by the blue-solid and orange-dashed lines in Fig.~\ref{two}. In principle, we can get the atomic excitation probability ${\left| {\varepsilon (t)} \right|^2}$ as a function of $\Gamma t$ with any $N$ in the above two cases. We take the number of coupling points $N = 6$ to exhibit typical features of the atomic excitation probability. Interestingly, the quantum interference effects between multiple coupling points and the semi-infinite waveguide lead to the approximate periodic oscillation behaviors of the dynamics, and the amplitudes of periodic oscillations do not decrease with time.

Similar to the case of the static bound state, we study the dependence of the field intensity function $P(x,t)$ on $x/x_0$ and $\Gamma t$ in Fig.~\ref{twoP32}. We show the time evolution of the field intensity function $P(x,t)$ at $x/x_0$ in Figs.~\ref{twoP32}(a) and \ref{twoP32}(c). The energy is bound between $0 \sim 6x_0$ with the position of the last coupling point as the boundary, while the energy outside the last coupling point disappears with time. The field intensity oscillates persistently after a long time due to the existence of bound states with two different frequencies in the waveguide. In Figs.~\ref{twoP32}(b) and \ref{twoP32}(d), we show the field intensity distribution calculated by Eq.~(\ref{phitime}) at $\Gamma t = 15$ corresponding to Figs.~\ref{twoP32}(a) and \ref{twoP32}(c).

In order to understand the influence of the number of coupling points on the atomic excitation probability, the dynamics with different number of coupling points ($N = 7,8$) in two cases given by Eqs.~(\ref{twoepsN}) and (\ref{twoepsN1}) is shown in Fig~\ref{twoN}. The amplitude and frequency of the atomic excitation probability change with the number of coupling points.

\begin{widetext}
\subsection{Periodic non-equal amplitude oscillating bound states}

We study the coexistence of bound states with three different frequencies ${\omega _{{k_0}}}$, ${\omega _{{k_1}}}$, and ${\omega _{{k_2}}}$ described by case (iii) in Sec.~\ref{si}. In this section, we analyze the situation that Eq.~(\ref{omegak1}) or (\ref{omegak2}) is satisfied when meeting Eq.~(\ref{omegak3}). The other cases of bound states with three frequencies are discussed in Appendix \ref{D}. With Eqs.~(\ref{sk12}) and (\ref{sk34}), the atomic dynamics in Eq.~(\ref{epst}) becomes
\begin{eqnarray}
\varepsilon (t) &=& \frac{1}{{1 + \frac{\Gamma }{2}\sum\limits_{m,n = 1}^N {\left[ { - \left| {m - n} \right|{\tau _0} + \left( {m + n} \right){\tau _0}} \right]} }}{e^{ - i2{k_0}\pi t/{\tau _0}}} \nonumber\\&&+ \frac{{2{{\sin }^2}({\omega _{{k_1}}}{\tau _0}/2)}}{{2{{\sin }^2}({\omega _{{k_1}}}{\tau _0}/2) + N\Gamma {\tau _0}}}{e^{ - i2{k_1}\pi t/\left( {N{\tau _0}} \right)}} + \frac{{2{{\sin }^2}({\omega _{{k_2}}}{\tau _0}/2)}}{{2{{\sin }^2}({\omega _{{k_2}}}{\tau _0}/2) + N\Gamma {\tau _0}}}{e^{ - i2{k_2}\pi t/\left( {N{\tau _0}} \right)}},
   \label{threeepsN}\\
\varepsilon (t) &=& \frac{1}{{1 + \frac{\Gamma }{2}\sum\limits_{m,n = 1}^N {\left[ { - \left| {m - n} \right|{\tau _0}{{\left( { - 1} \right)}^{\left| {m - n} \right|}} + \left( {m + n} \right){\tau _0}{{\left( { - 1} \right)}^{\left( {m + n} \right)}}} \right]} }}{e^{ - i(2{k_0} + 1)\pi t/{\tau _0}}} \nonumber\\&&+ \frac{{2{{\sin }^2}({\omega _{{k_1}}}{\tau _0}/2)}}{{2{{\sin }^2}({\omega _{{k_1}}}{\tau _0}/2) + N\Gamma {\tau _0}}}{e^{ - i2{k_1}\pi t/\left( {N{\tau _0}} \right)}} + \frac{{2{{\sin }^2}({\omega _{{k_2}}}{\tau _0}/2)}}{{2{{\sin }^2}({\omega _{{k_2}}}{\tau _0}/2) + N\Gamma {\tau _0}}}{e^{ - i2{k_2}\pi t/\left( {N{\tau _0}} \right)}},
   \label{threeepsN1}
\end{eqnarray}
\end{widetext}
where ${\omega_{k_n}}(n = 1, 2)$ in Eqs.~(\ref{threeepsN}) and (\ref{threeepsN1}) are given by Eqs.~(\ref{C4}) and (\ref{C5}), respectively. The derivation details of ${\omega_{k_n}}$ can be found in Appendix \ref{C}.
\begin{figure}[t]
\centering\scalebox{0.33}{\includegraphics{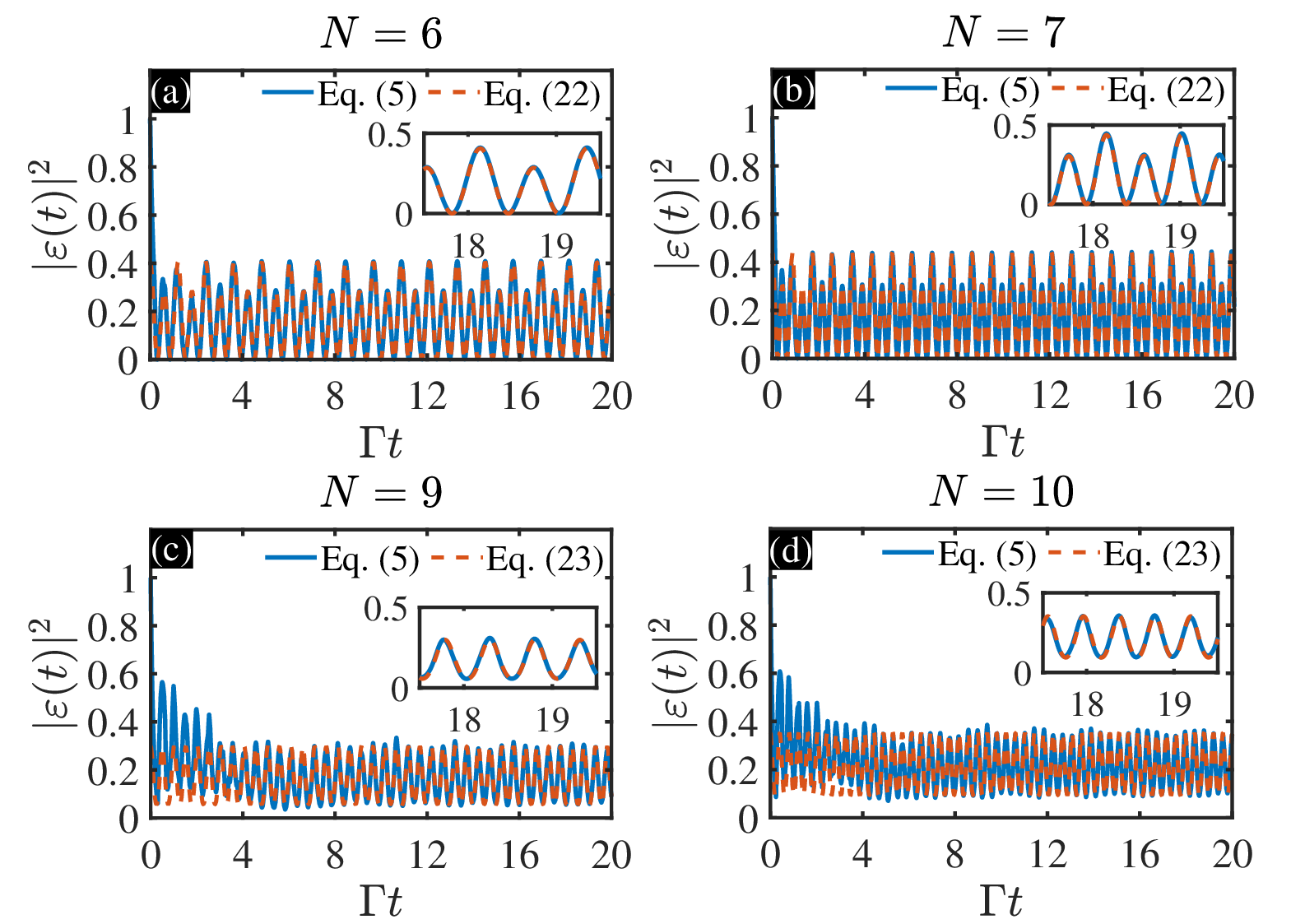}}
\caption{The influence of different number of coupling points $N$ on the non-equal amplitude oscillating bound states for the giant atom. The orange-dashed and blue-solid lines respectively correspond to the analytical solutions in Eqs.~(\ref{threeepsN}) and (\ref{threeepsN1}) and the numerical simulation of Eq.~(\ref{epsdottime}). The parameters chosen are (a) $N=6, k_{0}=4, k_{1} = 23, k_{2} = 25, \omega_0\tau_0=8\pi, \Gamma\tau_0=0.0642\pi$; (b) $N=7, k_{0}=4, k_{1} = 27, k_{2} = 29, \omega_0\tau_0=8\pi, \Gamma\tau_0=0.0393\pi$; (c) $N=9, k_{0}=8, k_{1} = 71, k_{2} = 82, \omega_0\tau_0=17\pi, \Gamma\tau_0=0.0989\pi$; (d) $N=10, k_{0}=8, k_{1} = 79, k_{2} = 91, \omega_0\tau_0=17\pi, \Gamma\tau_0=0.078\pi$. $\omega_{k_n}$ in (a)-(d) are determined by Eq.~(\ref{sk3}).}
\label{threeN}
\end{figure}
The dynamics with three coexisting bound states ${\omega _{{k_0}}}$, ${\omega _{{k_1}}}$, and ${\omega _{{k_2}}}$ in Eqs.~(\ref{threeepsN}) and (\ref{threeepsN1}) stems from the interaction of the photon through a semi-infinite waveguide and multiple coupling points, which is completely different from that in Refs.~\cite{Guo0430142020,Lim0237162023}, where the coexistence of at most two bound modes is achieved. The dynamics of the atomic excitation probability amplitude $\varepsilon (t)$ determined by Eqs.~(\ref{threeepsN}) and (\ref{threeepsN1}) in a semi-infinite waveguide is the superposition of three bound states with different frequencies in the long-time limit after the disappearance of all dissipative modes.

Figure \ref{three} shows the excitation dynamics of the giant atom, where the blue-solid and orange-dashed lines denote the numerical simulation of Eq.~(\ref{epsdottime}) and analytical results based on Eqs.~(\ref{threeepsN}) and (\ref{threeepsN1}), respectively. We find that atomic excitation dynamics of bound states with three frequencies exhibits non-equal amplitude persistent oscillation, where the differences between the dynamical characteristics in Figs.~\ref{three}(a) and \ref{three}(b) are explained in Appendix \ref{D}.
We use the field intensity function $P(x,t)$ as a function of $\Gamma t$ and $x/x_0$ and show the field intensity distribution in Fig.~\ref{threeP32}. Consistent with the static and equal amplitude oscillating bound states, the field intensity is bound between the mirror and the last coupling point. For clearer observation, we plot the field intensity distribution in the waveguide with the fixed parameter $\Gamma t = 20$ corresponding to Figs.~\ref{threeP32}(a) and \ref{threeP32}(c), as shown in Figs.~\ref{threeP32}(b) and \ref{threeP32}(d). We find that the field intensity distribution outside the last coupling point is $0$, which means that a perfect bound state is formed in the waveguide.

\begin{figure*}[t]
\centering\scalebox{0.445}{\includegraphics{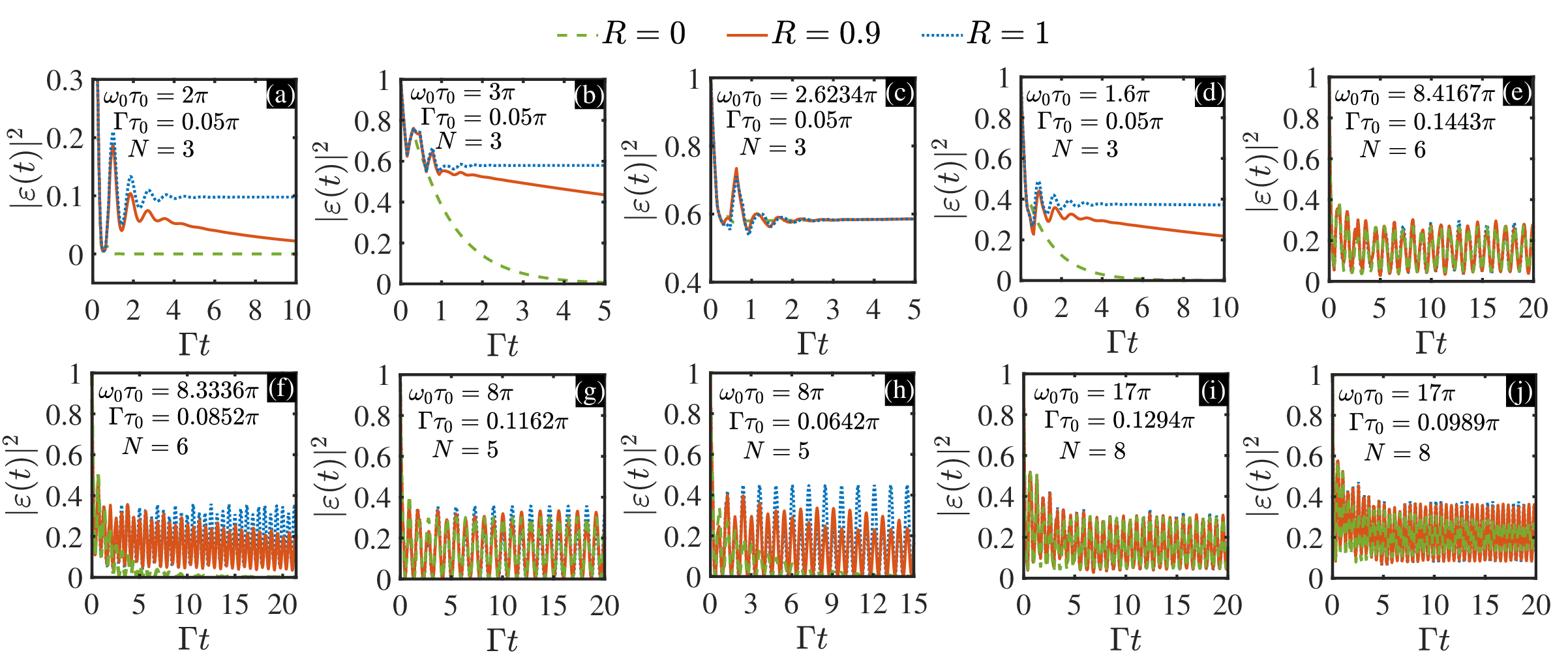}}
\caption{In order to understand the influence of reflectivity $R$ on the bound states, we plot the atomic excitation probability with different reflectivity $R$. The blue-dotted line denotes the ideal case, where the mirror has a perfect reflectivity $R=1$. The green-dashed and orange-solid lines respectively correspond to the case of no mirror $(R=0)$ and general case $(R=0.9)$. The atomic excitation probability ${\left| {\varepsilon (t)} \right|^2}$ is plotted by solving Eq.~(\ref{epsgamma}). The static bound states, equal amplitude oscillating bound states and non-equal amplitude oscillating bound states correspond to (a)-(d), (e)-(f), and (g)-(h), respectively. The parameters chosen are $\Gamma_{ext} = 0$ and $\lambda(t) = 0$ without considering the influence of the dephasing process and external decay.}
\label{mirror}
\end{figure*}

Finally, we plot the dynamics of the giant atom with different numbers of coupling points in Fig.~\ref{threeN}. The probability of the giant atom in the excited state depends on the number of coupling points. Tuning the number of coupling points can change the frequency and amplitude of atomic excitation probability.

\section{extension to general model}
\label{liu}

We focus on the non-Markovian dynamics of a two-level giant atom in a semi-infinite waveguide, where the waveguide is terminated by a perfect mirror with reflectivity $R=1$ at $x=0$. Considering the realistic experimental environment, we make some changes to the model to explain the detrimental factors.

Firstly, for the waveguide in the experiment, the reflectivity $R$ of the mirror is usually less than $1$. The complex probability amplitude $-r$ for backward reflection of the mirror satisfies ${| r |^2} = R$. Solving the one-dimensional scattering problem yields $r = R + i\sqrt {R(1 - R)} $. The delay term in Eq.~(\ref{epsdottime}) suggests the replacement $\frac{\Gamma }{2}\sum_{m,n = 1}^N {\varepsilon \left[ {t - (m + n){\tau _0}} \right]\Theta \left[ {t - (m + n){\tau _0}} \right]}  \to \frac{\Gamma }{2}r\sum_{m,n = 1}^N {\varepsilon \left[ {t - (m + n){\tau _0}} \right]\Theta \left[ {t - (m + n){\tau _0}} \right]} $ since the giant atom will reinteract only with the reflected part of the light.
In addition to the emissions of the atomic excited state into waveguide modes at rate ${\Gamma}$, we allow for an extra atomic coupling to a reservoir of external nonaccessible modes at a rate ${\Gamma _{ext}}$. When the external mode does not decay to the waveguide mode, the total effective dissipation rate of the excited atom is ${\Gamma _{tot}} = \Gamma  + {\Gamma _{ext}}$. It simply amounts to adding a term $ - {{N{\Gamma _{ext}}}}\varepsilon (t)/2$ on the right-hand side of Eq.~(\ref{epsdottime}).
Finally, we add a small white-noise stochastic term $\omega (t) = {\omega _0} + \lambda (t)$ to the excited state frequency to introduce (inhomogeneous) phase noise on the giant atom. Here the Gaussian distributed stochastic term $\lambda (t)$ is characterized by its zero mean ${\left\langle {\lambda (t)} \right\rangle _t} = 0$ and the noise correlation function $\left\langle {\lambda (t)\lambda (t')} \right\rangle  = 2\delta \omega \delta (t - t')$ ($\left\langle {...} \right\rangle $ stands for the ensemble average) \cite{Chenu1404032017,Khripkov0216062011}, where $\delta\omega$ denotes the associated dephasing rate.
\begin{figure*}
\centering\scalebox{0.45}{\includegraphics{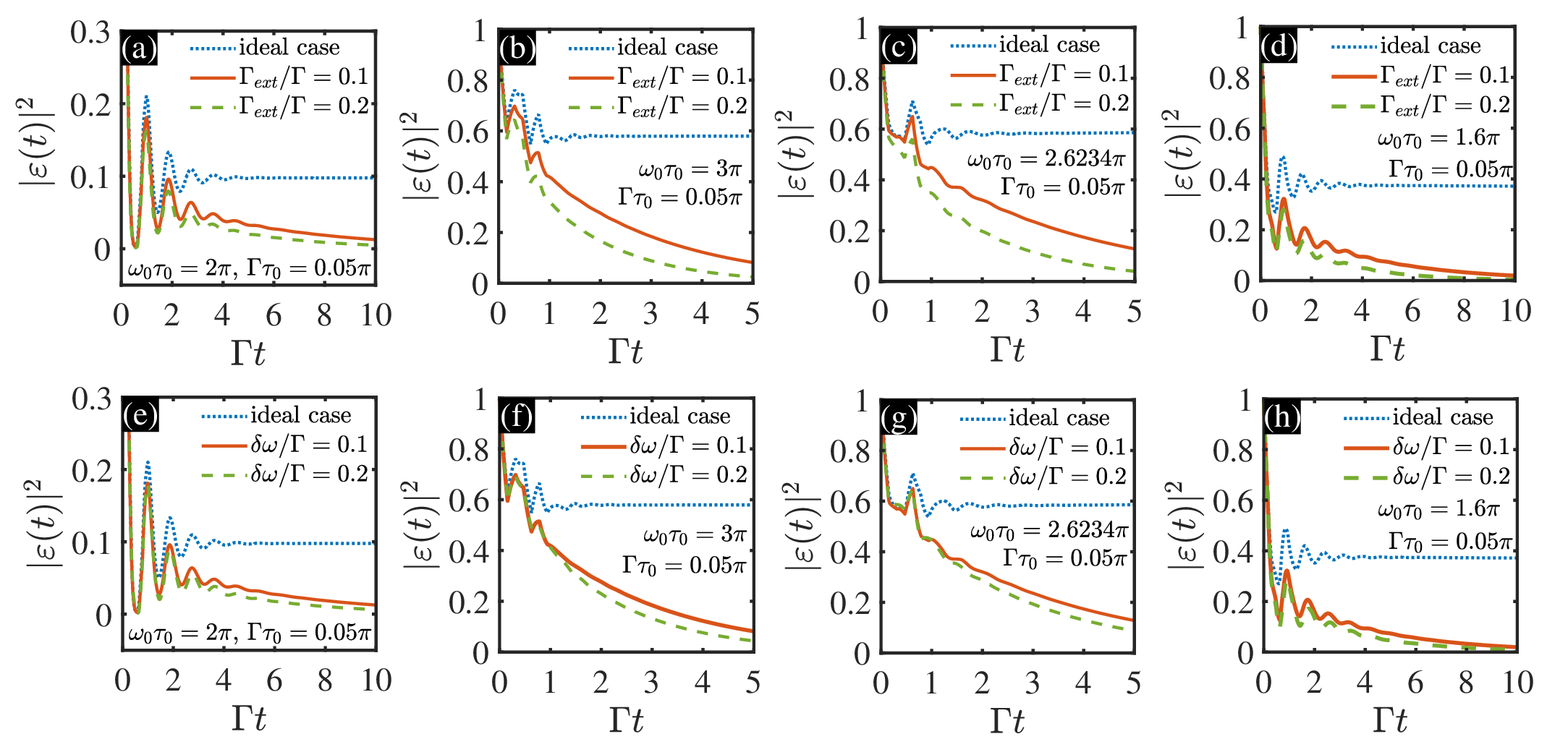}}
\caption{The influence of the external reservoir decay rate $\Gamma_{ext}$ and dephasing rate $\delta\omega$ on the static bound states with the number of coupling points $N = 3$. The blue-dotted line represents the ideal case ($R=1, \delta\omega=0, \Gamma_{ext}=0$). For (a)-(d), we fix $R=0.98$ and $\delta\omega=0.1\Gamma$ and vary the ratio between $\Gamma_{ext}$ and $\Gamma$ with keeping $\Gamma$ fixed. For (e)-(h), we remain $R=0.98$ and $\Gamma_{ext}=0.1\Gamma$, and change the dephasing rate $\delta\omega$.}
\label{influence1}
\end{figure*}
\begin{figure}
\centering\scalebox{0.47}{\includegraphics{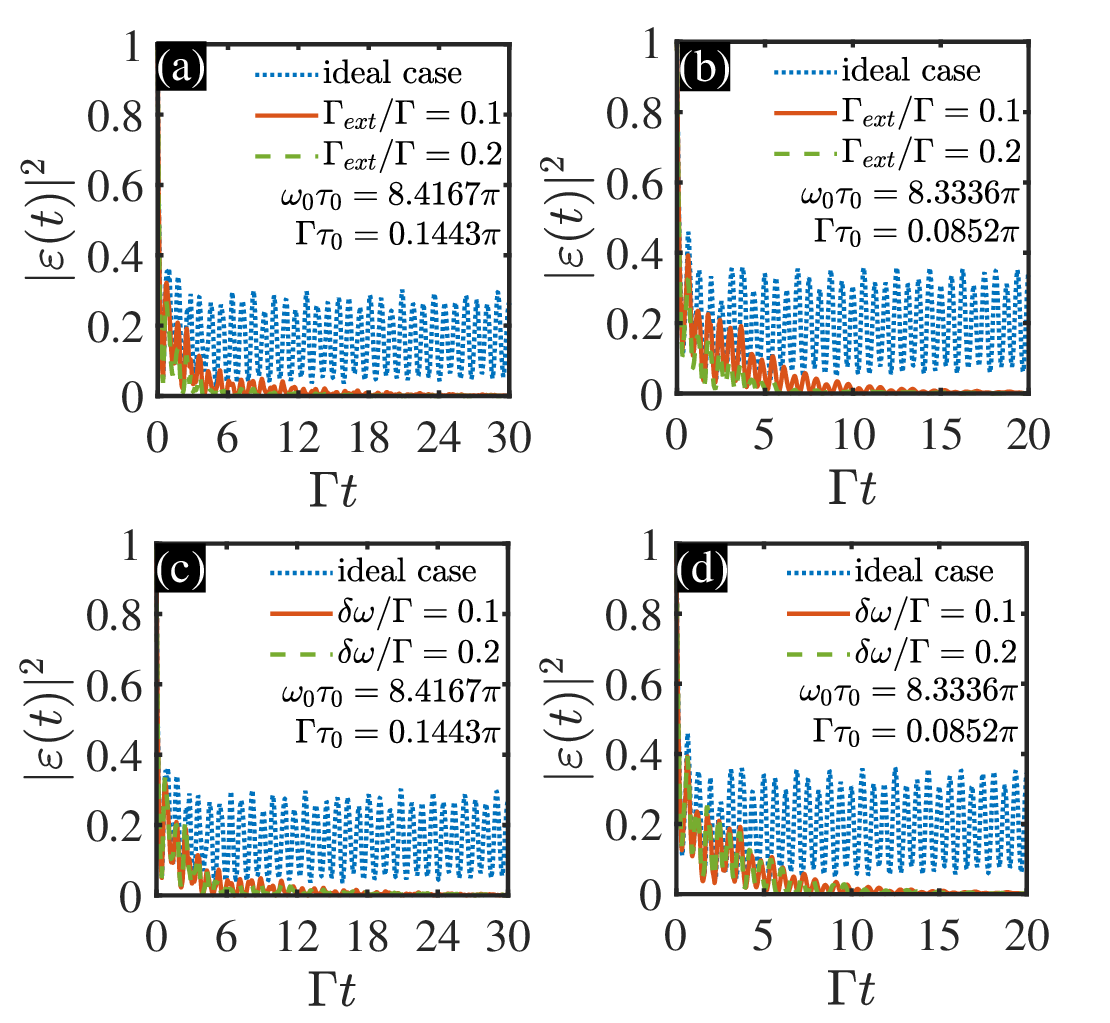}}
\caption{Influences of the different dephasing processes and external decay on the equal amplitude oscillating bound states for a giant atom with the number of coupling points $N = 6$. The atomic excitation probability ${\left| {\varepsilon (t)} \right|^2}$ is plotted by solving Eq.~(\ref{epsgamma}). (a)-(b) show the influences of different  external reservoir decay rates $\Gamma_{ext}$ on the atomic excitation probability with the fixed parameter $R = 0.98$ and $\delta\omega=0.1\Gamma$. The orange-solid and green-dashed lines respectively correspond to $\Gamma_{ext}=0.1\Gamma$ and $\Gamma_{ext}=0.2\Gamma$. In (c)-(d), we take $R = 0.98$ and $\Gamma_{ext}=0.1\Gamma$, where the orange-solid and green-dashed lines correspond to $\delta\omega=0.1\Gamma$ and $\delta\omega=0.2\Gamma$, respectively. The blue-dotted line corresponds to the ideal case with $R = 1$, $\Gamma_{ext}=0$, and $\delta\omega=0$.}
\label{influence2}
\end{figure}

In conclusion, Eq.~(\ref{epsdottime}) involved in the dephasing processes becomes
\begin{equation}
\begin{aligned}
\dot \varepsilon (t) =  &- i\left[{\omega _0} + \lambda (t)\right]\varepsilon (t) - \frac{{N{\Gamma _{ext}}}}{2}\varepsilon (t) \\&- \frac{\Gamma }{2}\sum\limits_{m,n = 1}^N {\varepsilon \left(t - \left| {m - n} \right|{\tau _0}\right)\Theta \left(t - \left| {m - n} \right|{\tau _0}\right)} \\&+ \frac{\Gamma }{2}r\sum\limits_{m,n = 1}^N {\varepsilon \left[ {t - \left(m + n\right){\tau _0}} \right]\Theta \left[ {t - \left(m + n\right){\tau _0}} \right]}.
   \label{epsgamma}
\end{aligned}
\end{equation}
When we adjust the reflectivity $R$ of the mirror at the end of the semi-infinite waveguide, the atomic excitation probabilities of the static, equal amplitude oscillating and non-equal amplitude oscillating bound states are shown in Fig.~\ref{mirror}. Except for Figs.~\ref{mirror}(c), \ref{mirror}(e), \ref{mirror}(g), and \ref{mirror}(i) obtained by Eqs.~(\ref{oneepsN}), (\ref{twoepsN}), (\ref{threeepsN}), and (\ref{threeepsN1}), the atomic excitation probabilities decay significantly when the reflectivity $R=0$, which denotes that the formation of the bound states except for these four cases strongly depend on the existence of the semi-infinite waveguide. The results got without considering the presence of a mirror $(R=0)$ are equivalent to those through a two-level giant atom interacting with a one-dimensional infinite waveguide.

\begin{figure*}[t]
\centering\scalebox{0.46}{\includegraphics{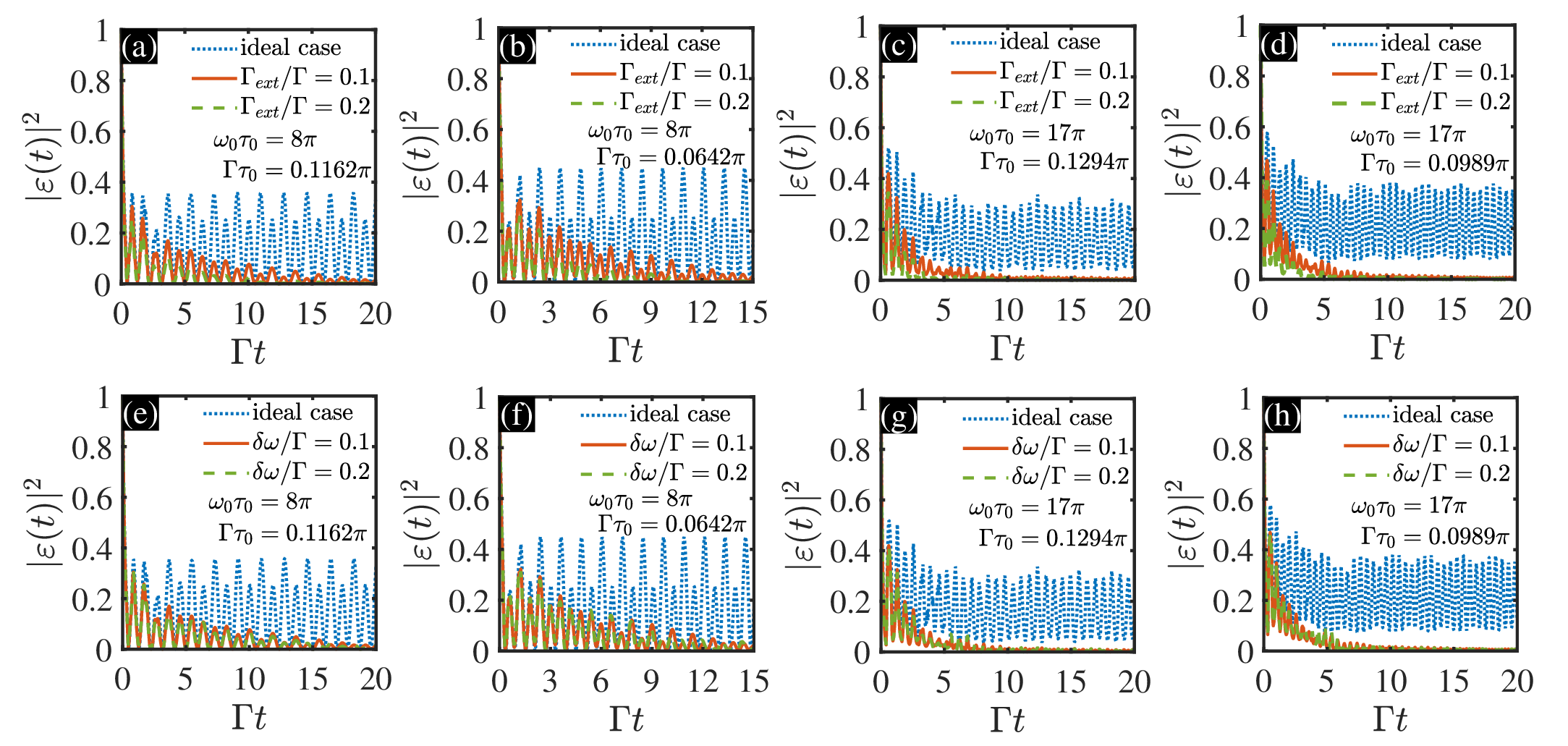}}
\caption{The figure shows the influences of the external reservoir decay rate $\Gamma_{ext}$ and dephasing rate $\delta\omega$ on the non-equal amplitude oscillating bound states. The blue-dotted line denotes the ideal case ($R=1, \delta\omega=0, \Gamma_{ext}=0$). The orange-solid and green-dashed lines respectively correspond to $\Gamma_{ext}=0.1\Gamma$ and $\Gamma_{ext}=0.2\Gamma$ in (a)-(d), while those in (e)-(h) respectively correspond to $\delta\omega=0.1\Gamma$ and $\delta\omega=0.2\Gamma$. The other parameters chosen are (a)-(d) $R=0.98, \delta\omega=0.1\Gamma$; (e)-(h) $R=0.98, \Gamma_{ext}=0.1\Gamma$. }
 \label{influence3},
\end{figure*}

For practical giant atom waveguide systems, in addition to dephasing, we also allow the giant atom to couple to a reservoir of external inaccessible modes and the semi-infinite waveguide of non-ideal mirrors. In Figs.~\ref{influence1}(a)-\ref{influence1}(d), we show the atomic excitation probabilities for static bound states with $R = 0.98$ and $\delta \omega  = 0.1\Gamma $. We give the different external reservoir decay rates ${\Gamma _{ext}} = \{ {0.1\Gamma ,0.2\Gamma } \}$ and compare them with the ideal case ($R = 1$, ${\Gamma _{ext}} = 0$, $\delta \omega  = 0$). The atomic probability decreases as the decay rate ${\Gamma _{ext}}$ increases, which indicates that the decay of the external inaccessible mode inhibits the formation of bound states. We plot the atomic probabilities with different dephasing rates $\delta \omega  = \{ {0.1\Gamma ,0.2\Gamma } \}$ and compare them with the ideal case ($R = 1$, ${\Gamma _{ext}} = 0$, $\delta \omega  = 0$) in Figs.~\ref{influence1}(e)-\ref{influence1}(h), where the other parameters chosen are $R = 0.98$ and ${\Gamma _{ext}} = 0.1\Gamma $. The results show  the inhibition effect of dephasing rate on probability, which is more obvious when the dephasing rate becomes larger. Figs.~\ref{influence2} and \ref{influence3} correspond to the influence of dephasing and external reservoir decay rates on the equal amplitude and non-equal amplitude oscillating bound states. Consistent with the static bound states, the atomic excitation probabilities decay rapidly when the dephasing and relaxation rates increase.

\section{multiple giant atoms are coupled to a one-dimensional semi-infinite waveguide}
\label{qi}
\begin{figure*}[t]
\centering\scalebox{0.5}{\includegraphics{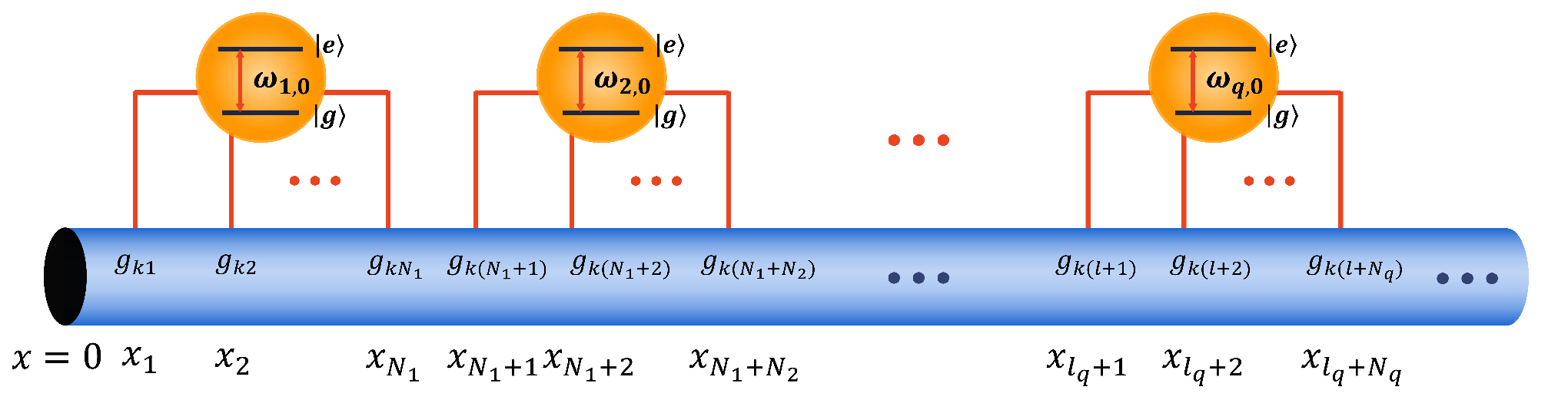}}
\caption{One-dimensional semi-infinite waveguide coupled to an array of noninteracting two-level giant atoms through $ \sum_{q = 1}^Q {{N_q}} $ coupling points with coupling coefficient $g_{km}$, where $N_q$ is the number of coupling points for the $q$th giant atom. The ground state $|g\rangle$ and the excited state $|e\rangle$ of the $q\mathrm{th}$ atom are separated in frequency by $\omega _{q,0}~(q = 1, 2,...,Q)$. $l_q$ denotes the sum of the number of all coupling points before the first coupling point of the $q$th giant atom, which is written as $l_q = \sum_{q'=1}^{q - 1} {{N_{q'}}} $. The distance between any two adjacent coupling points is $x_0$, which is equal to that between the mirror and the first coupling point.}\label{setupN}
\end{figure*}

We generalize the above results to a more general model involving $Q$ noninteracting two-level giant atoms coupled to a one-dimensional semi-infinite waveguide, where ground state $\left| g \right\rangle $ and excited state $\left| e \right\rangle $ of the $q$th atom are separated in frequency by $\omega _{q,0}~(q = 1, 2,...,Q)$. In Fig.~\ref{setupN}, $Q$ giant atoms interact with a one-dimensional semi-infinite waveguide through $\sum_{q = 1}^Q {{N_q}} $ coupling points, where $N_q$ denotes the number of coupling points for the $q$th giant atom. The distance between any adjacent coupling points is the same as that between the mirror and the first coupling point, both of which are $x_0$. $l_q$ denotes the sum of the number of all coupling points before the first coupling point of the $q$th giant atom, which is expressed as $l_q = \sum_{q'=1}^{q - 1} {{N_{q'}}} $. The total Hamiltonian of the system in the rotating-wave approximation is
\begin{equation}
\begin{aligned}
  {\hat{H}_Q} = &\sum\limits_{q = 1}^Q {{\omega _{q,0}}} {| e \rangle _{qq}}\langle e | + \int_0^{{k_c}} {dk} {\Omega _k}\hat a_k^\dag {{\hat a}_k} \\& +\sum\limits_{q = 1}^Q {\sum\limits_{m = l_q + 1}^{l_q + {N_q}} {\int_0^{{k_c}} {dk} {({g_{km}}\hat a_k^\dag {\sigma _q}_ -  + {g_{km}^*}{{\hat a}_k}{\sigma _q}_ + )}} },
\end{aligned}
\label{Hn}
\end{equation}
where ${\sigma _q}_+$ and ${\sigma _q}_ - $ are raising and lowering operators of the $q$th giant atom. The first and second terms of Eq.~(\ref{Hn}) describe the free Hamiltonian of $Q$ giant atoms and the one-dimensional semi-infinite waveguide, where the frequency of the $q$th atom is denoted by ${\omega _{q,0}} = {\omega _0} + {\delta _q}$. The last term of Eq.~(\ref{Hn}) results from the interaction between the $q$th giant atom and semi-infinite waveguide. We assume that the initial state is prepared in $| {\psi (0)} \rangle  = \sum_{q = 1}^Q {{\varepsilon _q}(0)} | {{e_q},0} \rangle $, where $| {{e_q},0} \rangle $ denotes that the $q$th giant atom is on the excited state $| e \rangle $ with the probability amplitude ${\varepsilon _q}(0)$, but simultaneously all the modes of the waveguide are all in the vacuum state. Schr\"odinger equation drives the initial state $| {\psi (0)} \rangle $ as
\begin{equation}
\left| {\psi (t)} \right\rangle  = \sum\limits_{q = 1}^Q {{\varepsilon _q}(t)} \left| {{e_q},0} \right\rangle  + \int {dk\varphi (k,t)} \hat a_k^\dag \left| {G,0} \right\rangle,
\label{psin}
\end{equation}
with $
 \left| {{e_q},0} \right\rangle  = {\left| g \right\rangle _1} \otimes {\left| g \right\rangle _2} \otimes  \cdots {\left| e \right\rangle _q} \otimes  \cdots {\left| g \right\rangle _Q} \otimes \left| 0 \right\rangle$ and $
 \left| {G,0} \right\rangle  = {\left| g \right\rangle _1} \otimes {\left| g \right\rangle _2} \otimes  \cdots {\left| g \right\rangle _q} \otimes  \cdots {\left| g \right\rangle _Q} \otimes \left| 0 \right\rangle.$
Substituting Eqs.~(\ref{Hn}) and (\ref{psin}) into Schr\"odinger equation, we obtain the probability amplitudes
\begin{eqnarray}
{{\dot \varepsilon }_q}(t) &=&  - i{\omega _{q,0}}{\varepsilon _q}(t) - i\sum\limits_{m = l_q + 1}^{l_q + {N_q}} {\int {dk{g_{km}^*}\varphi (k,t)} } ,
\label{dotepsilonn}\\
\dot \varphi (k,t) &=&  - i{\Omega _k}\varphi (k,t) - i\sum\limits_{q = 1}^Q {\sum\limits_{m = l_q + 1}^{l_q + {N_q}} {g_{km}{\varepsilon _q}(t)} } .
\label{dotphin}
\end{eqnarray}
Integrating Eq.~(\ref{dotphin}) with coupling coefficient ${g_{km}} = \sqrt {\Gamma v/\pi } \sin (km{x_0})$ results in
\begin{small}
\begin{equation}
\begin{aligned}
\varphi (k,t) =  \frac{1}{i}\int_0^t {\sum\limits_{q = 1}^Q {\sum\limits_{m = l_q + 1}^{l_q + {N_q}} {\sqrt {\frac{{\Gamma v}}{\pi }} \sin (km{x_0}){\varepsilon _q}(s)} } {e^{ - i{\Omega _k}\left( {t - s} \right)}}ds}.
\label{phin}
\end{aligned}
\end{equation}
\end{small}
With Eq.~(\ref{phin}), the probability amplitude in Eq.~(\ref{dotepsilonn}) becomes
\begin{small}
\begin{equation}
\begin{aligned}
\dot{\varepsilon}_q(t) = &- i\omega _{q,0}\varepsilon _q(t) - i\delta _q\varepsilon _q(t)\\ &- \frac{\Gamma }{2}\sum_{q = 1}^Q\sum_{m,n = l_q + 1}^{l_q + {N_q}} \varepsilon _q(t - |m - n|\tau _0)\\ &\times\Theta (t - |m - n|\tau _0)\\ &+ \frac{\Gamma }{2}\sum_{q = 1}^Q\sum_{m,n = l_q + 1}^{l_q + {N_q}} \varepsilon _q[t -(m + n)\tau _0]\\ &\times\Theta[t -(m + n)\tau _0].
\label{dotepsilonnt}
\end{aligned}
\end{equation}
\end{small}
Solving the set of time-delay differential equations in Eq.~(\ref{dotepsilonnt}), we can get complete information about the $Q$ noninteracting two-level giant atoms. Interactions between any two or more giant atoms caused by the semi-infinite waveguide generally exist, which leads to quantum correlations between different giant atoms. The result of Eq.~(\ref{dotepsilonnt}) is reduced to that in Eq.~(\ref{epsdottime}) when $Q = 1$. It can be predicted that when $Q\neq1$, there will be more modes of bound states in the waveguide simultaneously, which provides a positive way for us to control the coupled giant atoms through engineering the structured environment.

\section{conclusions and discussions}
\label{ba}
In this paper, we have investigated the non-Markovian dynamics in the spontaneous emission of a two-level giant atom interacting with a one-dimensional semi-infinite waveguide. We derived the analytical solutions for the atomic probability amplitude by Laplace transform, which show non-exponential dissipations due to the photon transferring between multiple coupling points and being reabsorbed after it is reflected by the semi-infinite waveguide. We discussed the conditions and origins for the formation of bound states and obtained three different bound states. According to the number of bound states in the waveguide, the bound states are divided into static case with one bound state, equal amplitude oscillation with two bound states, and non-equal amplitude oscillation with three bound states. The equal and non-equal amplitude oscillating bound states show the period oscillation behavior for the system, which can arise from the quantum interference effects between the multiple coupling points and semi-infinite waveguide. The oscillating bound states with multiple bound states provide a way for us to store and manipulate more complex quantum information. We assessed that the formation of bound states can be restricted in the presence of dissipation into unwanted modes and dephasing of the giant atom. Finally, we further expanded our analysis to a more general quantum system containing many noninteracting giant atoms interacting with a semi-infinite waveguide through multiple coupling points.

The study of non-Markovian dynamics in two-level giant atoms coupled to a one-dimensional semi-infinite waveguide may open up a way to better understand non-Markovian quantum networks and quantum communications. Moreover, compared to previous models, the obtained set of delay differential equations for the giant atom might pave the way to better understand the non-Markovian dynamics of many giant atoms coupled to a semi-infinite waveguide. As a prospect, we can further study anisotropic non-rotating wave two-level giant atomic systems through applying the method in Refs.~\cite{Shi1536022018,Lo0638072018,Shen0237072022,Shen0238562018,Xie0210462014,Chen0437082021,Rodriguez0466082008,Nakajima43631955,Frohlich8451950,Frohlich2911952,Lu0543022007} and driven three-level giant atomic systems in rotating-wave approximation, which are respectively described by
\begin{small}
\begin{equation}
\begin{aligned}
{{\hat H}_{2LS}} =& {\omega _e}| e \rangle \langle e| + \int_0^{{k_c}} {dk} {\Omega _k}\hat a_k^\dag {{\hat a}_k} + \sum\limits_m {\int_0^{{k_c}} {dk} } \\
&\cdot[{a_{km}}( {\hat a_k^\dag {\sigma _ - } + {{\hat a}_k}{\sigma _ + }} ) + {b_{km}}( {\hat a_k^\dag {\sigma _ + } + {{\hat a}_k}{\sigma _ - }} )] ,\\
{{\hat H}_{3LS}} =& {\omega _x}| x \rangle \langle x| + {\omega _e}| e \rangle \langle e| + \int_0^{{k_c}} {dk} \\
 &\cdot {\Omega _k}\hat a_k^\dag {{\hat a}_k} + \tilde \Omega_{3LS} {e^{i{\omega _l}t}}{\sigma _{ex}} + \tilde \Omega_{3LS} {e^{ - i{\omega _l}t}}{\sigma _{xe}}\\
& + \sum\limits_{n } {\int_0^{{k_c}} {dk} {g_{kn}}(\hat a_k^\dag {\sigma _{gx}} + {{\hat a}_k}{\sigma _{xg}})} ,
\label{threelevel}
\end{aligned}
\end{equation}
\end{small}
where $a_{km}$ and $b_{km}$ denote the coupling strengths of the rotating-wave and non-rotating-wave interactions, respectively. The transition from level $|e\rangle$ to $|x\rangle$ in Eq.~(\ref{threelevel}) is driven by the classical field with driving strength $\tilde \Omega_{3LS}$ and driving frequency $\omega_l$. Eq.~(\ref{threelevel}) will provide a way for us to further understand the influence of non-rotating wave effect and driving field on the dynamical evolution of giant atoms.

\section*{ACKNOWLEDGMENTS}

This work was supported by National Natural Science Foundation of China under Grant No. 12274064 and Natural Science Foundation of Jilin Province (subject arrangement project) under Grant No. 20210101406JC.

\appendix
\begin{widetext}
\section{Derivation of delay differential equations}
\label{A}
Integrating Eq.~(\ref{phidot}), we get
\begin{equation}
\varphi (k,t) = \varphi (k,0){e^{ - i\left[ {{\omega _0} + v(k - {k_0})} \right]t}} - i{e^{ - i\left[ {{\omega _0} + v(k - {k_0})} \right]t}}\int_0^t \varepsilon (s){\sum\limits_{m = 1}^N {\sqrt {\frac{{\Gamma v}}{\pi }} } \sin (km{x_0}){e^{  i\left[ {{\omega _0} + v(k - {k_0})} \right]s}}ds}.
\label{A1}
\end{equation}
With $\varepsilon (0) = 1$ and $\varphi (k,0) = 0$, substituting Eq.~(\ref{A1}) into Eq.~(\ref{epsdot}) gives
\begin{equation}
\dot \varepsilon (t) =  - i{\omega _0}\varepsilon (t) - \frac{{\Gamma v}}{\pi }\int_0^t {\varepsilon (s){e^{i{\omega _0}(s - t)}}{e^{ - iv{k_0}(s - t)}}ds} \int_0^{{k_c}} {\sum\limits_{m,n = 1}^N {\sin (km{x_0})\sin (kn{x_0}){e^{ivk(s - t)}}dk} } .
\label{A2}
\end{equation}
Through $\int_{0}^{k_c}dk\rightarrow\int_{-\infty}^{+\infty}dk$ \cite{Roy0210012017,Shen3020012005,Zhang0323352020,Liao0630042016}, Eq.~(\ref{A2}) can be rewritten as
\begin{equation}
\begin{aligned}
\dot \varepsilon (t) =  &- i{\omega _0}\varepsilon (t) - \frac{\Gamma }{{4\pi }}\int_0^t \varepsilon  (s){e^{i{\omega _0}(s - t)}}{e^{ - iv{k_0}(s - t)}}ds\int_{ - \infty }^{ + \infty } {\sum\limits_{m,n = 1}^N {\left\{ {{e^{ivk\left[ {s - \left( {t - \left( {m - n} \right){\tau _0}} \right)} \right]}} + {e^{ivk\left[ {s - \left( {t - \left( {n - m} \right){\tau _0}} \right)} \right]}}} \right.} }  \\&+ {e^{ivk\left[ {s - \left( {t - \left( {m + n} \right){\tau _0}} \right)} \right]}}\left. { + {e^{ivk\left[ {s - \left( {t + \left( {m + n} \right){\tau _0}} \right)} \right]}}} \right\}d(vk),
\label{A3}
\end{aligned}
\end{equation}
where $\tau_0=x_0/v$ denotes the delay time between two coupling points. Eq.~(\ref{A3}) is reduced to
\begin{equation}
\begin{aligned}
\dot \varepsilon (t) =  &- i{\omega _0}\varepsilon (t) - \frac{\Gamma }{2}\sum\limits_{m,n = 1}^N \varepsilon  \left[ {t - (m - n){\tau _0}} \right]\Theta \left[ {t - (m - n){\tau _0}} \right] - \frac{\Gamma }{2}\sum\limits_{m,n = 1}^N \varepsilon  \left[ {t - (n - m){\tau _0}} \right]\Theta \left[ {t - (n - m){\tau _0}} \right]{\text{ }} \\ &+ \frac{\Gamma }{2}\sum\limits_{m,n = 1}^N \varepsilon  \left[ {t - (m + n){\tau _0}} \right]\Theta \left[ {t - (m + n){\tau _0}} \right] + \frac{\Gamma }{2}\sum\limits_{m,n = 1}^N \varepsilon  \left[ {t + (m + n){\tau _0}} \right]\Theta \left[ {t + (m + n){\tau _0}} \right],
\label{A4}
\end{aligned}
\end{equation}
where we have used the identities $\int_{a-\epsilon}^{a+\epsilon}f(x)\delta(x-a)dx=f(a)$, $\frac{1}{{2\pi }}\int_{ - \infty }^{ + \infty } {{e^{ik(x - a)}}dk = \delta (x - a)} $, and the dispersion relation $\omega_0=vk_0$. Eq.~(\ref{epsdottime}) can be obtained due to the last term disappearing of Eq.~(\ref{A4}).
The total time-dependent field function in the waveguide
\begin{equation}
\phi (x,t) = \sqrt {\frac{2}{\pi }} \int {\varphi (k,t)\sin (kx)dk}=- i\sqrt {\frac{2}{\pi }} \int {{e^{ - ivkt}}\int_0^t { \varepsilon (s)\sum\limits_{m = 1}^N {\sqrt {\frac{{\Gamma v}}{\pi }} \sin (km{x_0})\sin (kx){e^{ivks}}dsdk} } }
\label{A5}
\end{equation}
leads to Eq.~(\ref{phitime}) by repeating the similar calculations with the above derivations.

\section{Derivation of Eq.~(\ref{real})}
\label{B}
Substituting $s_k=-i\omega_k$ into Eq.~(\ref{sk}) and expanding the sum terms
\begin{equation}
\begin{aligned}
\sum\limits_{m,n = 1}^N {{e^{i|m - n|{\omega _k}{\tau _0}}}}  &= 2\frac{{N - (N + 1){e^{i{\omega _k}{\tau _0}}} + {e^{i(N + 1){\omega _k}{\tau _0}}}}}{{{{\left( {1 - {e^{i{\omega _k}{\tau _0}}}} \right)}^2}}} - N,\\
\sum\limits_{m,n = 1}^N {{e^{i(m + n){\omega _k}{\tau _0}}}}  &= \frac{{{e^{i 2{\omega _k}{\tau _0}}}{{\left( {1 - {e^{iN{\omega _k}{\tau _0}}}} \right)}^2}}}{{{{\left( {1 - {e^{i{\omega _k}{\tau _0}}}} \right)}^2}}},
\label{B1}
\end{aligned}
\end{equation}
we obtain
\begin{equation}
i(\omega_0-\omega _k)+\frac{\Gamma}{2}\left[2\frac{N-(N+1)e^{i\omega_k\tau_0}+e^{i(N+1)\omega_k\tau_0}}{\left(1-e^{i\omega_k\tau_0}\right)^2}
-\frac{e^{i 2\omega_k\tau_0}\left(1-e^{iN\omega_k\tau_0}\right)^2}{\left(1-e^{i\omega_k\tau_0}\right)^2}-N\right]=0.
\label{B2}
\end{equation}
The two expressions in Eq.~(\ref{real}) correspond to the real and imaginary parts of Eq.~(\ref{B2}), respectively.
\end{widetext}

\begin{figure}
\centering\scalebox{0.34}{\includegraphics{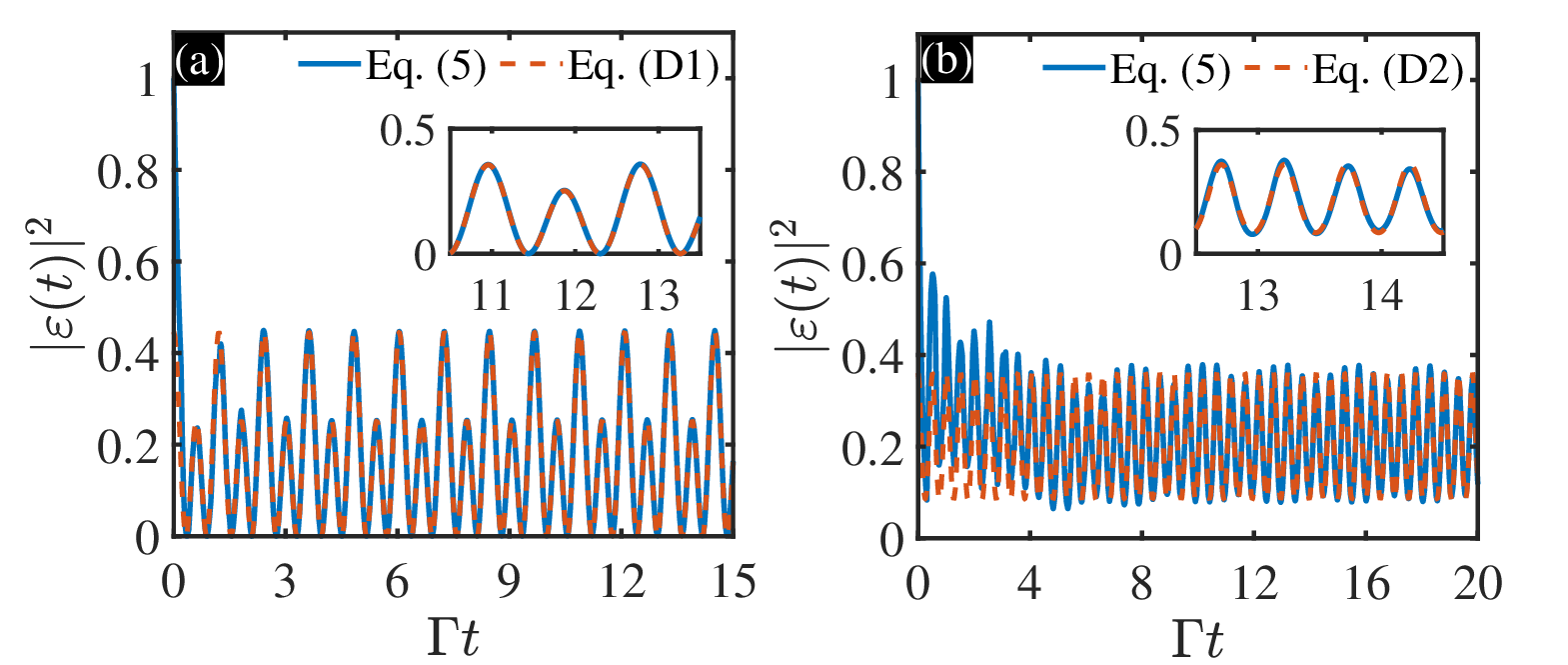}}
\caption{The non-equal amplitude oscillating bound states of a giant atom. The orange-dashed and blue-solid lines correspond to the analytical solutions in  Eqs.~(\ref{D1}) and (\ref{D2}) and numerical simulation with Eq.~(\ref{epsdottime}), respectively. The parameters chosen are (a) $N=5, k_{0}=4, k_{1} = 23, k_{2} = 25, \omega_0\tau_0=8\pi, \Gamma\tau_0=0.0642\pi$; (b) $N=8, k_{0}=8, k_{1} = 71, k_{2} = 82, \omega_0\tau_0=17\pi, \Gamma\tau_0=0.0989\pi$. $\omega_{k_n}$ in (a) and (b) are determined by Eq.~(\ref{sk4}).}
\label{appendix1}
\end{figure}
\begin{figure}[H]
\centering\scalebox{0.338}{\includegraphics{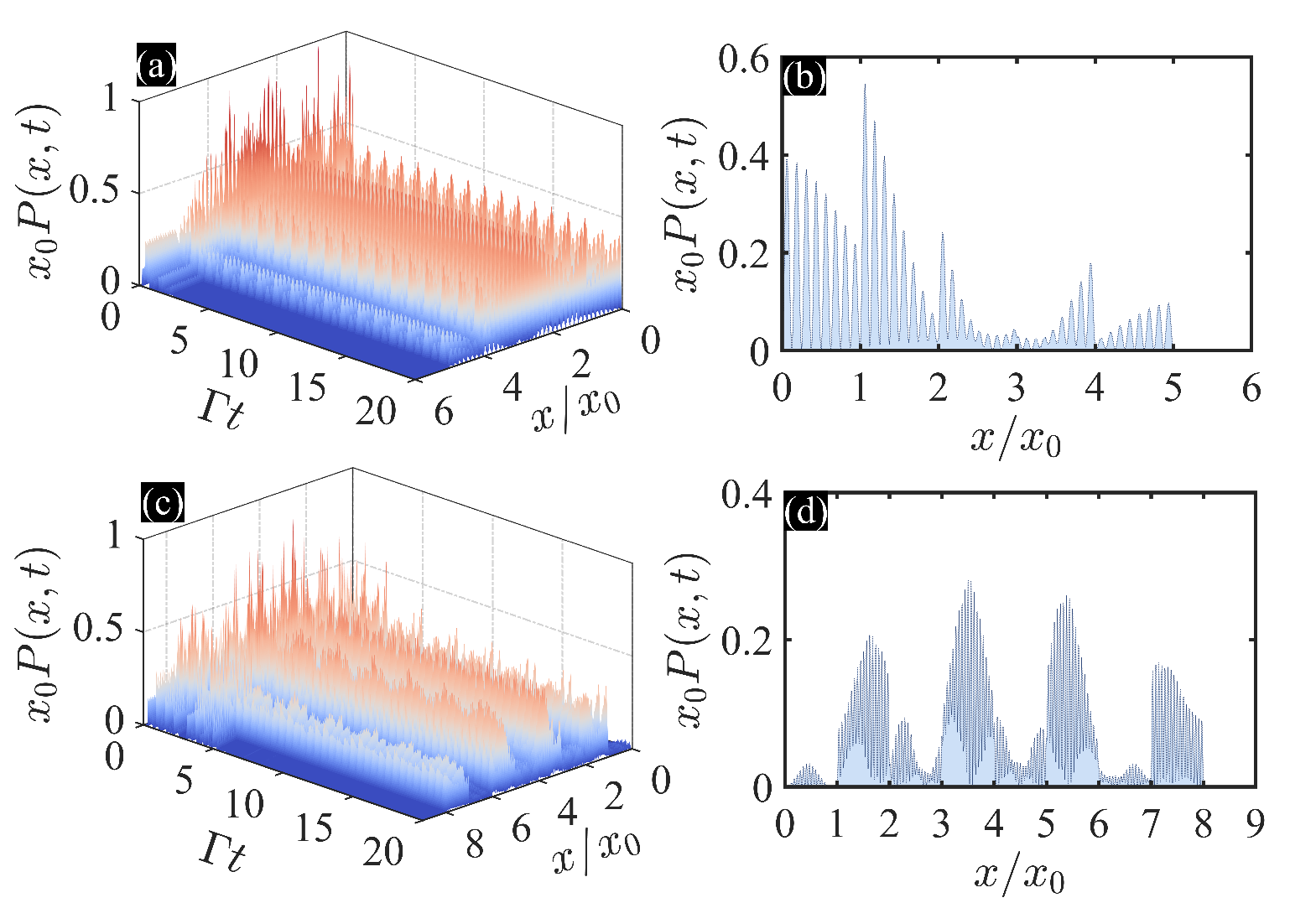}}
\caption{The field intensity $P(x,t) = {\left| {\phi (x,t)} \right|^2}$ is plotted as a function of $\Gamma t$ and $x/x_0$ based on Eq.~(\ref{phitime}). (a) and (c) show the time evolution of the field intensity $P(x,t)$ in the waveguide correspond to Figs.~\ref{appendix1}(a) and \ref{appendix1}(b), where the parameters $k_0, k_1,$ and $k_2$ are the same as those in Fig.~\ref{appendix1}. The colors in (a) and (c) describe the field intensity of the bound states. For clarity, (b) and (d) show the distribution of $P(x,t)$ in the waveguide at $t=20/\Gamma$, which correspond to (a) and (c). The blue-dotted line is the numerical simulation of Eq.~(\ref{phitime}). The other parameters are the same as those in Fig.~\ref{appendix1}.}
\label{appendix2}
\end{figure}

\section{Discussion on the coexistence of multiple bound states}
\label{C}
When $k_1$ and $k_2$ are simultaneously the two solutions of Eq.~(\ref{omegak3}), the parameters $\omega\tau_0$ and $\Gamma\tau_0$ have to be
\begin{equation}
\begin{aligned}
  \omega_0\tau_0=&\frac{2k_1\pi}{N}-\frac{2(k_1-k_2)\pi}{N}\frac{\cot\left(\frac{k_1\pi}{N}\right)}{\cot\left(\frac{k_1\pi}{N}\right)-\cot\left(\frac{k_2\pi}{N}\right)}>0,\\
  \Gamma\tau_0=&\frac{4(k_1-k_2)\pi}{N^2}\frac{1}{\cot\left(\frac{k_1\pi}{N}\right)-\cot\left(\frac{k_2\pi}{N}\right)}>0.
  \label{C1}
\end{aligned}
\end{equation}
If both the bound state conditions (\ref{omegak1}) and (\ref{omegak3}) are satisfied, three bound states coexist in the system. Substituting ${\omega _0}{\tau _0} = {\omega _{{k_0}}}{\tau _0} = 2{k_0}\pi$ into Eq.~(\ref{omegak3}) leads to $\Gamma {\tau _0} = \left( {\frac{{2{k_1}\pi }}{N} - 2{k_0}\pi } \right) \frac{2}{N}\tan \left( {\frac{{{k_1}\pi }}{N}} \right) = \frac{{4\pi }}{{{N^2}}}\tan \left( {\frac{{{k_1}\pi }}{N}} \right)\left( {{k_1} - {k_0}N} \right)$, which is reduced to $
  \Gamma {\tau _0} =q \frac{{4\pi }}{{{N^2}}}\tan \left( {\frac{{q\pi }}{N}} \right)$ by introducing ${k_1} = N{k_0} + q$,
where ${q,~~{k_0},~~{k_1} \in {{\Bbb N}_ + }}$. We get ${k_2} = N{k_0} - q$ due to the tangent function being an odd function. Collecting all these together, the frequencies in Eq.~(\ref{threeepsN}) read
\begin{equation}
\begin{aligned}
{\omega _{{k_0}}}{\tau _0} &= 2{k_0}\pi, \\
{\omega _{{k_1}}}{\tau _0} &= \frac{{2{k_1}\pi }}{N} = 2{k_0}\pi  + \frac{{2q\pi }}{N},\\
{\omega _{{k_2}}}{\tau _0} &= \frac{{2{k_2}\pi }}{N} = 2{k_0}\pi  - \frac{{2q\pi }}{N}.
\label{C4}
\end{aligned}
\end{equation}
With the same method, we obtain the frequencies in Eq.~(\ref{threeepsN1}) when Eqs.~(\ref{omegak2}) and (\ref{omegak3}) exist together
\begin{equation}
\begin{aligned}
  {\omega _{{k_0}}}{\tau _0} &= (2{k_0} + 1)\pi, \\
  {\omega _{{k_1}}}{\tau _0} &= \frac{{2{k_1}\pi }}{N} = 2{k_0}\pi  - \frac{{2q\pi }}{N}, \\
  {\omega _{{k_2}}}{\tau _0} &= \frac{{2{k_2}\pi }}{N} = (2{k_0} + 1)\pi  + \frac{{2q\pi }}{N}.
  \label{C5}
\end{aligned}
\end{equation}
Replacing $N$ with $N+1$, a series of results can be got when Eq.~(\ref{omegak4}) holds or both Eqs.~(\ref{omegak12}) and (\ref{omegak4}) are satisfied.

\begin{widetext}
\section{The non-equal amplitude oscillating bound states caused by the coexistence of Eqs.~(\ref{omegak12}) and (\ref{omegak4})}
\label{D}
Here, we focus on the coexistence of three mode bound states when Eqs.~(\ref{omegak12}) and (\ref{omegak4}) are met simultaneously, which results in that Eqs.~(\ref{threeepsN}) and (\ref{threeepsN1}) become
\begin{eqnarray}
\varepsilon (t) &=& \frac{1}{{1 + \frac{\Gamma }{2}\sum\limits_{m,n = 1}^N {\left[ { - \left| {m - n} \right|{\tau _0} + \left( {m + n} \right){\tau _0}} \right]} }}{e^{ - i2{k_0}\pi t/{\tau _0}}} \nonumber\\&&+ \frac{{2{{\sin }^2}({\omega _{{k_1}}}{\tau _0}/2)}}{{2{{\sin }^2}({\omega _{{k_1}}}{\tau _0}/2) + (N + 1)\Gamma {\tau _0}}}{e^{ - i2{k_1}\pi t/\left[ {(N + 1){\tau _0}} \right]}} + \frac{{2{{\sin }^2}({\omega _{{k_2}}}{\tau _0}/2)}}{{2{{\sin }^2}({\omega _{{k_2}}}{\tau _0}/2) + (N + 1)\Gamma {\tau _0}}}{e^{ - i2{k_2}\pi t/\left[ {(N + 1){\tau _0}} \right]}},
   \label{D1}\\
\varepsilon (t) &=& \frac{1}{{1 + \frac{\Gamma }{2}\sum\limits_{m,n = 1}^N {\left[ { - \left| {m - n} \right|{\tau _0}{{\left( { - 1} \right)}^{\left| {m - n} \right|}} + \left( {m + n} \right){\tau _0}{{\left( { - 1} \right)}^{\left( {m + n} \right)}}} \right]} }}{e^{ - i(2{k_0} + 1)\pi t/{\tau _0}}} \nonumber\\&&+ \frac{{2{{\sin }^2}({\omega _{{k_1}}}{\tau _0}/2)}}{{2{{\sin }^2}({\omega _{{k_1}}}{\tau _0}/2) + (N + 1)\Gamma {\tau _0}}}{e^{ - i2{k_1}\pi t/\left[ {(N + 1){\tau _0}} \right]}} + \frac{{2{{\sin }^2}({\omega _{{k_2}}}{\tau _0}/2)}}{{2{{\sin }^2}({\omega _{{k_2}}}{\tau _0}/2) + (N + 1)\Gamma {\tau _0}}}{e^{ - i2{k_2}\pi t/\left[ {(N + 1){\tau _0}} \right]}}.\nonumber\\
 \label{D2}
\end{eqnarray}
We take the atomic excitation probability as a function of ${\Gamma t}$ in Fig.~\ref{appendix1}. The excitation probability of the giant atom is a non-equal amplitude oscillating bound state, which is consistent with the case when both Eqs.~(\ref{omegak12}) and (\ref{omegak3}) are met. In Fig.~\ref{appendix2}, we plot the corresponding time evolution of the field intensity in the waveguide and the field intensity distribution at $t=20/\Gamma$.
We note that there is a clear difference in frequency and amplitude between Figs.~\ref{appendix1}(a) and \ref{appendix1}(b), which is similar to the phenomenon between Figs.~\ref{three}(a) and \ref{three}(b), although they are both superpositions of three different bound modes. In order to understand the reason for the difference, we start with the expression of the atomic excitation probability amplitude. When case (iii) in Sec.~\ref{si} is satisfied, the long-time dynamics of the giant atom can be written as
\begin{equation}
\varepsilon (t) = {A_0}{e^{ - i{\omega _{{k_0}}}t/{\tau _0}}} + {A_1}{e^{ - i{\omega _{{k_1}}}t/{\tau _0}}} + {A_2}{e^{ - i{\omega _{{k_2}}}t/{\tau _0}}},
 \label{D3}
\end{equation}
which results in
\begin{equation}
\begin{aligned}
{\left| {\varepsilon (t)} \right|^2} = &{\left| {{A_0}} \right|^2} + {\left| {{A_1}} \right|^2} + {\left| {{A_2}} \right|^2} \\&+ 2{A_0}{A_1}\cos \left[ {\left( {{\omega _{{k_0}}} - {\omega _{{k_1}}}} \right){\tau _0}t} \right] + 2{A_0}{A_2}\cos \left[ {\left( {{\omega _{{k_0}}} - {\omega _{{k_2}}}} \right){\tau _0}t} \right] + 2{A_1}{A_2}\cos \left[ {\left( {{\omega _{{k_1}}} - {\omega _{{k_2}}}} \right){\tau _0}t} \right],
  \label{D4}
\end{aligned}
\end{equation}
where
\begin{equation}
\begin{aligned}
{A_0} = \frac{1}{{1 + \frac{\Gamma }{2}\sum\limits_{m,n = 1}^N {\left[ { - \left| {m - n} \right|{\tau _0} + \left( {m + n} \right){\tau _0}} \right]} }},~~
{A_\alpha } = \frac{{2{{\sin }^2}({\omega _{{k_\alpha }}}{\tau _0}/2)}}{{2{{\sin }^2}({\omega _{{k_\alpha }}}{\tau _0}/2) + N\Gamma {\tau _0}}},
  \label{D5}
\end{aligned}
\end{equation}
with $\alpha=1,2$. Together with Eq.~(\ref{C4}), we get $| {{\omega _{{k_0}}}{\tau _0} - {\omega _{{k_1}}}{\tau _0}} | = | {{\omega _{{k_0}}}{\tau _0} - {\omega _{{k_2}}}{\tau _0}} | = 2q\pi {\tau _0}/N,| {{\omega _{{k_1}}}{\tau _0} - {\omega _{{k_2}}}{\tau _0}} | = 4q\pi {\tau _0}/N$, and substituting them into Eq.~(\ref{D5}) leads to
\begin{equation}
\begin{aligned}
{\left| {\varepsilon (t)} \right|^2} = &{\left| {{A_0}} \right|^2} + {\left| {{A_1}} \right|^2} + {\left| {{A_2}} \right|^2} + 2{A_0}\left( {{A_1} + {A_2}} \right)\cos (\Upsilon {\tau _0}t) + 2{A_1}{A_2}\cos (2\Upsilon {\tau _0}t),
 \label{D7}
\end{aligned}
 \end{equation}
with $\Upsilon  = 2q\pi /N$.
Therefore, the non-equal amplitude oscillating bound state is a superposition of the cosine functions of two persistent oscillations with  amplitudes $2{A_0}\left( {{A_1} + {A_2}} \right)$ and $2{A_1}{A_2}$, respectively. We define the difference between the amplitudes of the two cosine functions as ${\Delta _A} = \left| {2{A_0}\left( {{A_1} + {A_2}} \right) - 2{A_1}A} \right|$. As shown in Figs.~\ref{three}(a) and \ref{appendix1}(a), the difference between two adjacent peaks is small when ${\Delta _A}$ is large. On the contrary, when ${\Delta _A}$ becomes small, the difference between two adjacent wave peaks becomes large as depicted in Figs.~\ref{appendix1}(b) and \ref{three}(b). This is the essential reason leading to the difference between Figs.~\ref{three}(a) and \ref{three}(b) (Figs.~\ref{appendix1}(a) and \ref{appendix1}(b)).

\end{widetext}

\end{document}